\documentclass[aps,twocolumn,prd,superscriptaddress]{revtex4-2}

\usepackage{dcolumn}
\usepackage{bm}
\usepackage{bbold}

\usepackage[utf8]{inputenc}
\usepackage{babel}

\usepackage{mathtools}
\usepackage{amsfonts}
\usepackage{mathrsfs}
\usepackage{bbm}
\usepackage{slashed}
\usepackage{tensor}

\usepackage{graphicx}
\usepackage{color}
\usepackage{array}
\usepackage[abs]{overpic}

\usepackage{placeins}

\usepackage{makecell}
\usepackage{caption}
\usepackage{subcaption}

\usepackage{xspace}
\usepackage{siunitx}
\usepackage{xfrac}
\usepackage{hyperref}
\usepackage[nameinlink]{cleveref}
\usepackage{appendix}
\usepackage{units}
\usepackage{xifthen}
\usepackage{xcolor}
\hypersetup{
	colorlinks,
	linkcolor={red!75!black},
	citecolor={blue!75!black},
	urlcolor={blue!75!black}
}

\usepackage{booktabs}
\usepackage{multirow}

\newcolumntype{C}{>{$}c<{$}}
\AtBeginDocument{
	\heavyrulewidth=.08em
	\lightrulewidth=.05em
	\cmidrulewidth=.03em
	\belowrulesep=.65ex
	\belowbottomsep=0pt
	\aboverulesep=.4ex
	\abovetopsep=0pt
	\cmidrulesep=\doublerulesep
	\cmidrulekern=.5em
	\defaultaddspace=.5em
}

\def\pslash{p\llap{/}}

\graphicspath{{./figures/}}

\newcommand{\gettitle}{Shocks and quark-meson scatterings at large density}

\newcommand{\getHeidelbergAffiliation}{\affiliation{Institut f{\"u}r Theoretische Physik, Universit{\"a}t Heidelberg, Philosophenweg 16, 69120 Heidelberg, Germany}}
\newcommand{\getEMMIAffiliation}{\affiliation{ExtreMe Matter Institute EMMI, GSI, Planckstr. 1, 64291 Darmstadt, Germany}}
\newcommand{\getStonyBrookAffiliation}{\affiliation{Center for Nuclear Theory, Department of Physics and Astronomy,
Stony Brook University, Stony Brook, New York 11794-3800, USA}}

\hypersetup{
	pdftitle={\gettitle},
	pdfauthor={Grossi,  Wink},
	pdfkeywords={discontinuous galerkin}
	{functional renormalization group} {effective potential}
	{phase transition} {numerical methods} {phase structure},
	bookmarksopen=true,
	bookmarksopenlevel=2,
	bookmarksnumbered=true
}

\begin{document}

\title{\gettitle}

\author{Eduardo Grossi}\getStonyBrookAffiliation
\author{Friederike J. Ihssen}\getHeidelbergAffiliation
\author{Jan M. Pawlowski}\getHeidelbergAffiliation\getEMMIAffiliation
\author{Nicolas Wink}\getHeidelbergAffiliation

\begin{abstract}
	We discuss the phase structure of the two-flavour quark-meson model including quantum, thermal, density and critical fluctuations with the functional renormalisation group. This study combines two technical advances in the literature, that are also chiefly important for the quantitative access of the phase boundary of QCD at large density or baryon chemical potential. Specifically we allow for the formation and propagation of shocks as well as a fully self-consistent computation of the order parameter potential for chiral symmetry breaking.   
\end{abstract}

\maketitle 

\section{Introduction}\label{sec:intr}

The theoretical access to the QCD phase structure at large densities is chiefly important for our understanding of running and planned heavy-ion experiments, for reviews see \cite{Luo:2017faz, Adamczyk:2017iwn, Andronic:2017pug, Stephanov:2007fk, Andersen:2014xxa,  Shuryak:2014zxa, Pawlowski:2014aha, Roberts:2000aa, Fischer:2018sdj, Yin:2018ejt}. At large densities functional approaches, both within QCD and in low energy effective models (LEFTs)  have made rapid progress within the past two decades, for results in the present context with the functional renormalisation group (fRG) see e.g.~\cite{Fu:2018qsk, Fu:2019hdw, Leonhardt:2019fua, Braun:2019aow, Braun:2020ada, Otto:2020hoz, Dupuis:2020fhh, Fu:2021oaw, Chen:2021iuo}, for results with Dyson-Schwinger equations (DSE) see e.g.~\cite{Roberts:2000aa, Fischer:2014ata, Gao:2016qkh, Fischer:2018sdj, Isserstedt:2019pgx, Gao:2020qsj, Gao:2020fbl}. At small densities these studies are accompanied by respective  lattice studies, see e.g.~\cite{Bazavov:2017dus, Bazavov:2017tot, Bonati:2018nut, Borsanyi:2018grb, Bazavov:2018mes, Guenther:2018flo, Ding:2019prx, Borsanyi:2020fev}. By now the results from both, lattice and functional methods, agree at small densities. In turn, at larger densities the lattice is hampered by the sign problem, while the approximations to the full QCD effective action within functional approaches require systematic qualitative improvements. 

Chiefly important are the introduction of a Fierz-complete basis of four-quark scattering vertices as well as quantitative access to order parameter potentials for homogeneous and inhomogeneous condensates. The latter allows us to discuss the eminently important question of the location of phase transition lines, that of the symmetry breaking pattern and the order of the phase transitions. It has been shown in the past decade that functional QCD flows towards QCD-assisted low energy effective models for energy scales below $1$\,GeV, for a detailed discussion see in particular the recent works \cite{Fu:2019hdw, Dupuis:2020fhh}.  With dynamical hadronisation \cite{Gies:2001nw, Gies:2002hq, Pawlowski:2005xe, Floerchinger:2009uf} the LEFT is the (Polyakov-loop enhanced) quark-meson model (QM-model), or more generally the quark-hadron model. For recent fRG-works with the (P)QM model on the phase structure of QCD revelant for the present work see e.g.\ \cite{Skokov:2010wb, Herbst:2010rf, Kamikado:2012cp,  Braun:2012zq, Tripolt:2013zfa, Herbst:2013ail, Herbst:2013ufa, Mitter:2013fxa, Drews:2013hha, Helmboldt:2014iya, Pawlowski:2014zaa, Fu:2015naa, Jiang:2015xqz, Fu:2016tey, Rennecke:2016tkm, Zhang:2017icm, Tripolt:2017zgc, Resch:2017vjs, Fu:2018wxq, Fu:2018swz, Yin:2019ebz, Wen:2019ruz, CamaraPereira:2020xla, Otto:2020hoz, Braun:2020bhy, Connelly:2020gwa},  for a recent overview see \cite{Dupuis:2020fhh}. This emergence of LEFTs from first principle QCD flows is well understood and quantitatively explored in the vacuum, see \cite{Gies:2002hq, Braun:2008pi, Braun:2014ata, Mitter:2014wpa, Cyrol:2017ewj, Alkofer:2018guy}. It entails that the infrared critical dynamics is dominated by the low energy fluctuations of quarks and hadrons. For small baryon-chemical potentials, $\mu_B/T\lesssim 4$, the relevant hadronic degrees of freedom are simply the pseudoscalar pions and the sigma mode, see \cite{Fischer:2018sdj, Fu:2019hdw, Braun:2019aow, Gao:2020fbl, Gao:2020qsj}. In turn, for baryon-chemical potentials $\mu_B\lesssim 4/T$ the situation is less clear, but we expect sizable diquark contributions, see \cite{Braun:2019aow}. 

In the present work we make significant steps towards such a quantitative control of the phase structure of high density QCD within functional methods. It combines two systematic advances in the past years: The first one was the development of self-consistent approximations for the computation of order parameter potentials,  \cite{Pawlowski:2014zaa}. The second one was the development of a numerical approach for solving flow equations that also enables us to discuss discontinuities in the flows such as shocks that are potentially relevant for the correct description of  first and second order phase transitions, \cite{Grossi:2019urj}. Within this approach we compute the phase structure of the quark-meson model (QM-model) at finite temperature and density. An important benchmark is already provided in the large $N_f$-limit with an infinite number of flavours. It is argued that within an 't Hooft-type limit we can mimic the two-flavour QM-model well (or any other flavour), and in particular reproduce well it's non-universal properties such as the location of the phase boundary. Moreover, in this limit the numerical approach with discontinuous Galerkin set-up in \cite{Grossi:2019urj} is fully developed and we have a quantitative access to the shock-development and propagation. The respective results are compare with the currently most advanced approximation (including shocks) to the self-consistent approximation including the order parameter potentials in \cite{Pawlowski:2014zaa} for the $N_f=2$-flavour quark-meson model. The results include also the regime $\mu_B/T\gtrsim 4$. In this regime the current model has to be augmented with a diquark channel which is done in a forthcoming work. Still, the present work is a necessary and important study also in this regime.

\section{Quark Meson Model}\label{sec:qm}
The quark-meson model describes the dynamics of quarks and mesons at low energy. Within functional QCD this low energy effective theory (LEFT) emerges naturally from the momentum scale flow of the theory at momentum or cutoff scales  $k \lesssim 1$\,GeV, \cite{Gies:2002hq, Braun:2008pi, Braun:2014ata, Mitter:2014wpa, Rennecke:2015eba, Cyrol:2017ewj, Alkofer:2018guy, Fu:2019hdw}. In this regime the gluonic degrees of freedom decouple from the dynamics due to the gluonic mass gap in QCD, for a detailed discussion  see \cite{ Cyrol:2017ewj, Fu:2019hdw, Dupuis:2020fhh}, for a discussion of the emergent LEFT see \cite{Alkofer:2018guy}. 

\subsection{Emergent LEFTs and their range of validity}
The key ingredient for this emergence is the scale-dependent four-quark scattering, whose dynamics at large momentum scales is driven by a box diagram with a two-gluon exchange between quark currents. For the discussion of its low-momentum behaviour we restrict ourselves to the momentum-independent tensor structures, that is 10 tensor structures in two-flavour QCD and 28 (32) tensor structures in three-flavour QCD, the relevant cases for the discussion of the phase structure of QCD. It has been shown in \cite{Mitter:2014wpa, Cyrol:2017ewj} that in the vacuum the scalar-pseudoscalar channel is dominating the dynamics by far, both above and below the chiral symmetry breaking scale of $k\approx 500$\,MeV: switching of all other channels leads to negligible effects for most physical observables. Moreover, in \cite{Braun:2019aow} is has been shown for $N_f=2$-flavour QCD in the chiral limit, that qualitatively this dominance persists up to large densities or chemical potentials, $\mu_B/T_c(0) \approx 6$, where $T_c(\mu_B)$ is the chiral crossover or phase transition temperature  at a given baryon chemical potential $\mu_B$. This highly interesting first dominance study in QCD is based on qualitative approximations, and a conservative error estimate leads us to $\mu_B/T_c(0) \lesssim 4-8$ for the (total) dominance regime of the scalar-pseudoscalar channel. 

This supports the computations in \cite{Fu:2019hdw}, where the phase structure of 2- and 2+1-flavour QCD was computed within a one-channel approximation (scalar-pseudoscalar) to the Fierz-complete tensor structure for $\mu_B/T(\mu_B)\lesssim 6$ or $\mu_B/T(\mu_B)\lesssim 4$. Then, dynamical hadronisation takes into account multi-scattering events of the resonant channels (multi-scatterings of pions and the scalar $\sigma$-mode) that are relevant for the critical dynamics in a regime with second or first order transitions. In summary we estimate the reliability regime of the present approximations in functional QCD (see also respective considerations in DSEs (\cite{Fischer:2018sdj, Gao:2020fbl, Gao:2020qsj}) to be 
\begin{align}\label{eq:mu/T-Estimate}
	\mu_B/T(\mu_B)\lesssim 4\,.
\end{align}
The critical end point (CEP) computed both within the most recent fRG-computations, $\mu_B/T(\mu_B)=5.59$ from \cite{Fu:2019hdw} and DSE, $\mu_B/T(\mu_B)=5.54$ from \cite{Gao:2020qsj}, for the physical case of 2+1-flavour QCD agree well, which sustains the respective reliability of these estimates. Still it is not within the regime of quantitative reliability of the current approximation. 

Consequently, \labelcref{eq:mu/T-Estimate} entails that for a quantitatively sound prediction of the CEP the current approximation to the full first principle QCD-flow has to be improved systematically in two directions for chemical potentials $\mu_B/T(\mu_B)\gtrsim 4$: First we need to include at least the dominant tensor structure at large densities, the csc- or diquark-channel. This extension will be considered elsewhere. Second the self-consistent computation of the order parameter potential set-up in \cite{Pawlowski:2014aha} is required. This is done in the present work within a recently developed numerical approach that also allows the inclusion of the formation and propagation of shocks, \cite{Grossi:2019urj}. 

\subsection{Quark-Meson Model with the functional renormalisation group} 

In this section we briefly recapitulate the fRG-approach to the (Polyakov-enhanced) Quark-Meson model (QM-model). The inclusion of the dynamical mesons as low energy effective degrees of freedom has to be seen as an efficient and convenient book-keeping device for the respective resonant interaction channels. In particular, this substitutes the rather tedious inclusion of the resonant parts of the higher-order scattering processes of quarks. Still, if used on a quantitative level, even for large UV-cutoff scales its effective action does not reduce to a simple local classical actions. For more details and in particular its quantitative properties as an emergent low energy theory in QCD we refer to  \cite{Mitter:2014wpa, Braun:2014ata, Cyrol:2017ewj, Fu:2019hdw}. Validity checks, benchmarks and bounds in comparison to QCD have been provided in \cite{Alkofer:2018guy}. 

As discussed before, in the present work we restrict ourselves to a globally rather qualitative approximation to the effective action. Here, we are predominantly concerned with the quantitative access to the effective potential of the chiral order parameter. The systematic inclusion of the present quantitative setup within functional QCD flows is straightforward due to the modular nature of the fRG-approach and will be considered elsewhere. 

The scales of the present LEFT are gauged by the pion decay constant in the chiral limit. We use $f_{\pi,\chi}=88$\,MeV and measure all other scales with these units.  

We choose the UV-cutoff scale of the QM-model as $\Lambda=650$\,MeV. We consider this to be a good compromise between integrating-out as many momentum-fluctuations as possible and stretching the validity-bound of the LEFT. The momentum fluctuations with momentum scales $k\leq \Lambda$ are included with the functional renormalisation group (fRG). This approach has been used intensively in the past 25 years for the inclusion of low energy dynamics of the QM-model. For the setup of the flow equation for the effective action, and the derivation of the respective flow equations for (field-dependent) couplings we refer to the fRG-reviews see \cite{Berges:2000ew, Schaefer:2006sr, Gies:2006wv, Braun:2011pp, Dupuis:2020fhh}. Applications relevant for the present work can be found in \cite{Schaefer:2004en,Pawlowski:2014zaa, Braun:2019aow}, the derivations and flows for the present approximation can be found in \cite{Pawlowski:2014zaa}. 

The effective action $\Gamma_k[q,\bar q, \phi]$ of the $N_f$-flavour QM-model is used in the following approximation, 
\begin{align}\nonumber 
\Gamma_k[q,\bar q, \phi]=&\,
\int_x \Bigl\{ \bar{q} (\gamma_\mu \partial_\mu - \gamma_0 \mu_q) q + \frac{1}{2}(\partial_\mu \phi)^2 \\[1ex]
&\hspace{0cm}
+ h_k(\rho) \,\bar{q} ( \tau_0 \sigma + \bm{\tau}  \bm{\pi} ) q + V_k(\rho) - c_\sigma \sigma \Bigr\} 
\,,
\label{eq:EffAct}\end{align}
with $\tau_\mu$ being related to the Pauli matrices, $\tau=1/2 (\mathbb{1}, i \gamma_5 \bm \sigma)$, and the quark-meson coupling incorporates the SU(2) $\cong$ SO(3) symmetry of the pseudoscalar subgroup.  The O(4)-scalar field $\phi$ and the respective O(4)-invariant $\rho$ are given by 
\begin{align}
	\phi = (\sigma, \vec{\pi})^T \,, \qquad \rho = \frac{1}{2}\phi^2 = \frac{1}{2}(\vec{\pi}^2 + \sigma^2) \,.
\end{align}
In \labelcref{eq:EffAct} we have also $\int_x = \int_{0}^{1/T}dx_0 \int d^3x $ as an abbreviation for the finite temperature spatial integration. 

We emphasize that the the Yuakwa coupling $h_k(\rho)$ is considered fully field-dependent. It multiplies the O(4)-invariant operator $\bar q\,\tau\phi\,q$, hence it only depends on the O(4)-invariant $\rho$. The field-dependence of the Yukawa-coupling takes into account higher-order point-like scatterings of the resonant scalar-pseudo-scalar channels with the quark--anti-quark pair. The inclusion of these processes is necessary for a fully consistent zeroth order derivative expansion, and has been introduced in \cite{Pawlowski:2014zaa}. For further works in Yukawa models with field-dependent Yukawa coupling see  \cite{Vacca:2015nta, Jakovac:2015kka, Gies:2017zwf, Yin:2019ebz, Fejos:2020lli}. This is easily seen by performing a perturbative one-loop computation within the QM-model. Then, the quark loop with $h(\rho)$ contributes to the full effective potential. Of course higher terms in the derivative expansion also contribute to the effective potential, but the Yukawa-term contains no derivatives. Accordingly, its full field-dependence should be accounted for in a consistent lowest order derivative expansion. 

Finally, the scalar effective potential $V_k(\rho)$ takes  into account the remaining part of the higher orders scatterings of the mesons. The linear term introduces explicit chiral symmetry breaking (finite current quark masses). Evidently, it drops out on the right hand side of the flow equation and does not run. Consequently, the full flow and hence the full effective potential $V_k$ does not know anything about explicit chiral symmetry breaking, and we do not consider it any further.  

The next systematic step beyond the zeroth order derivative expansion would be the inclusion of wave-function renormalisations $Z_q(\phi), Z_\phi(\phi)$ for quarks and mesons. This can be done either fully field-dependent (1st order derivative expansion) or field-independent (usually called LPA'). The latter approximation has been used in \cite{Pawlowski:2014zaa} together with a field-dependent Yukawa-coupling. While technically in reach, we have chosen to drop these terms for the sake of concentrating on the quantitative discussion of the full effective potential. Hence, with \labelcref{eq:EffAct} we assume implicitly, 
\begin{align} \label{eq:z}
Z_{q,k}(\rho) =1 =  Z_{\phi,k}(\rho)\,.
\end{align}

The flow equation for the complete set of couplings, $h_k(\rho), V_k(\rho)$, and wave function renormalisations, can be found in \cite{Pawlowski:2014zaa}. We use the same setup here, including the choice of regulators, three-dimensional flat or Litim regulators, \cite{Litim:2002cf}. 

For the effective potential we simply evaluate the flow for $\Gamma_k[q,\bar q,\phi]$ for constant scalar fields $\phi$ and vanishing quark fields, $q,\bar q=0$. This leads us to 
 \begin{align}\nonumber 
 \partial_t V_k(\rho)=& \frac{k^5}{12 \pi} \Biggl[- \frac{ 4 N_f N_c}{\epsilon_k^q}\Bigl(1  -n_f(\epsilon_k^q + \mu)-n_f(\epsilon_k^q - \mu)\Bigr)\\[1ex]
& \hspace{-1cm}+\frac{N_f^2 -1}{\epsilon_k^\pi} \Bigl(1 + 2n_B(\epsilon_k^\pi)\Bigr)+\frac{1}{\epsilon_k^\sigma} \Bigl(1 + 2n_B(\epsilon_k^\sigma)\Bigr) 
\Biggr]\,,  	
  \label{eq:udrhoqm} 
 \end{align}
 with $\rho$-dependent quark- and meson-dispersion relations, 
 \begin{align} \label{eq:disp}
 \epsilon_k^\phi(\rho) = \sqrt{k^2 + m_{\phi,k}^2(\rho)}\,,\quad \epsilon_k^q(\rho) = \sqrt{k^2 + m_{q,k}^2(\rho)}
 \,, 
 \end{align} 
and the RG-time $t=\ln k/\Lambda$. The RG-time involves a reference scale in the logarithm, which we have set to be the initial scale. The  masses $m_q, m_\phi$ are obtained by evaluating the respective two-point functions at constant fields. Note that $m_q, m_\phi$ are the curvature and not the pole or screening masses of quarks and mesons, for respective definitions and discussions see \cite{Helmboldt:2014iya}. 

The meson curvature masses are defined with 
\begin{align} \nonumber 
m_{\pi,k}^2(\rho) =&\, \partial_\rho V_k(\rho) \,, \\[1ex] 
m_{\sigma,k}^2(\rho) =&  \partial_\rho V_k(\rho)  + 2 \rho \partial_\rho^2 V_k(\rho) \,,
\label{eq:mphi}\end{align} 
and hence are curvature-coefficients of the effective potentials. In turn, the quark mass is proportional to the Yukawa-coupling, 
\begin{align} \label{eq:mpsi}
m_{q,k}^2(\rho) = 2h_k(\rho)^2\rho
\,.
\end{align}
It is left to discuss the flow equation for the field-dependent Yukawa coupling, for details we again refer  to \cite{Pawlowski:2014zaa}. We can project the flow for $\Gamma_k$ onto the Yukawa coupling $h(\rho)$ by  evaluating the quark two-point function at vanishing quark and pion fields, $q,\bar q, \bm \pi =0$, and constant $\sigma$. With \labelcref{eq:EffAct} we arrive at 
\begin{align} \nonumber 
\Gamma^{(2)}_{q\bar q,k}[\sigma](p) \delta_{p,p'}=&\, \left. 
\frac{\delta^2 \Gamma[q,\bar q, \phi]}{\delta q(p) \delta \bar q(p')} \right|_{q,\bar q, \bm \pi=0}\\[1ex]
\simeq  &\, i \pslash - \gamma_0 \mu  
+\frac 12 h_k(\rho)  \sigma  - c_\sigma \sigma\,,  
 \label{eq:quark-2point}\end{align}
where we have dropped the momentum conservation $\delta_{p,p'}$ in the last line with $\delta_{p,p'}=(2 \pi)^4\delta (p-p')$ in the vacuum. 

\begin{figure}[t!]
	\centering
	\includegraphics[width=1\linewidth]{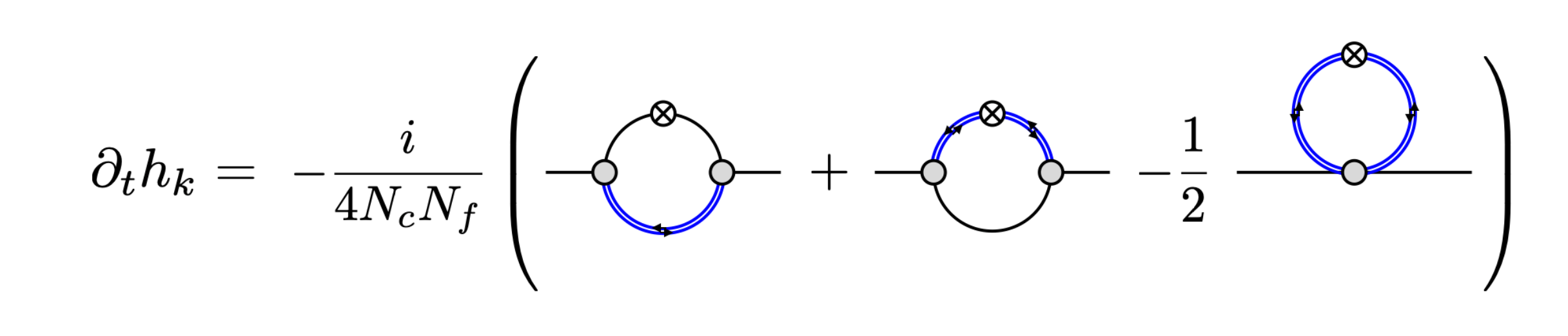}
	\caption{Diagrammatic representation of the flow of the Yukawa coupling. The circled cross $\otimes$ depicts the regulator insertion $\partial_t R_k(\vec p^2)$ and the gray dots denote full vertices. The double lines depict the mesons and indicate the quark content. The arrows depict the quark number flow. \hspace*{\fill} }
	\label{fig:h}
\end{figure}
\labelcref{eq:quark-2point} reflects the fact that the Yukawa term simply is the $\rho$-dependent mass quark term,  $m_{q,k}(\sigma)=h_k(\rho)\sigma/2$.  Accordingly, the flow of the Yukawa coupling $h_k(\rho)$ can be derived from that of the scalar part of the quark two-point function: it is simply $\sigma/2 \,\partial_t h_k(\rho)$ as $\partial_t c_\sigma=0$ by definition. Thus we get, 
\begin{align}
\label{eq:yuk}
\partial_t h_k(\rho) = - \frac{1}{4 N_c N_f}\frac{1}{\sigma}\mathrm{Re}\mathrm{Tr} \,  \Gamma^{(2)}_{q\bar q,k}[\sigma](p=0)\,.
\end{align} 
In \labelcref{eq:yuk} we have used that $\rho=\sigma^2/2$ for $\bm \pi=0$. The flow \labelcref{eq:yuk} is depicted in \Cref{fig:h}.
From \labelcref{eq:yuk} and the approximation \labelcref{eq:EffAct}  we finally get, 
\begin{align} \nonumber
\partial_t h_k(\rho)=&\,
4 v_3 h_k^3(\rho) \big[L^{(4)}_{(1,1)}(m^2_{q,k},m^2_{\sigma,k};T,\mu)\\[1ex] \nonumber
&
- (N_f^2-1) L^{(4)}_{(1,1)}(m^2_{q,k},m^2_{\pi,k};T,\mu)\big]\\[1ex] \nonumber
& + 16 v_3 h_k(\rho)h'_k(\rho) \rho \big[h_k(\rho) + \rho h'_k(\rho)\big] \\[1ex] \nonumber
& \times L^{(4)}_{(1,1)}(m^2_{q,k},m^2_{\sigma,k};T,\mu) \\[1ex]  \nonumber
&-2 v_3 k^2 \big[(3 h'_k(\rho) +2 \rho h''_k(\rho))l_1^{(B,4)}(m^2_{\sigma,k};T)\\[1ex]
& + (N_f^2 -1) h'_k(\rho) l_1^{(B,4)}(m^2_{\pi,k};T)  \big] \,.
\label{eq:hdrhoqm}
\end{align}
with 
\begin{align} 
v_{d-1}=\frac{1}{2^{d+1}\pi^{\frac{d}{2}}\Gamma\left(\frac{d}{2}\right)}\,.
\end{align}
The threshold functions $l_1^{(B,4)}$ originate in the bosonic loops in 4 dimensions. The $L^{(4)}_{(1,1)}$ originate in the mixed fermionic and bosonic contributions, again in 4 dimensions. Both functions are defined in \Cref{app:thfkt}. The the first and second lines from \labelcref{eq:hdrhoqm} are contributions of the first two diagrams with mixed fermionic and bosonic loops in \Cref{fig:h}, whereas the third and fourth lines are that from the bosonic loop with the four-vertex.

\subsection{The Large-N Limit}\label{sec:largeN}

Most of the numerical results in the present work are obtained in the large-N limit of these equations, as it simplifies the numerical treatment significantly:  it eliminates the $\sigma$-loop in the flow equation, and hence the second derivative terms in the sigma meson mass term. We are left with only the pion loops as well as the quark loop. The pion loops constitute the flow of a pure $O(N)$-theory in the large $N$-limit as considered in \cite{Grossi:2019urj} with discontinuous Galerkin methods. Such a non-linear first order system is solved using approximate Riemann solvers. These solvers rely on the assumption that the solution is dominated by one strong wave, for more details see \Cref{app:numerics}. This assumption holds if the flow is dominated by contributions of the pion and quark loops, which is always ensured in the large-N limit.

This simplification is also helpful when considering systems of multiple differential equations.  However, it will also be instructive to make a comparison between both the finite $N_f$ case and the large $N_f$ limit in the case with constant Yukawa coupling. Moreover, we can simulate the physics case,  $N_f=2$ and $N_c=3$, with a suitable chosen large $N$-limit: 

To begin with we keep the ratio of color and flavour fixed to that of the $N_f=2$ quark-meson model in QCD, 
\begin{subequations}
	\label{eq:rescale}
\begin{align}\label{eq:cf}
	\frac{N_c}{N_f} = \frac{3}{2} \,. 
\end{align}
With \labelcref{eq:cf} we keep the flavour-colour balance of QCD intact. This ensures that the contributions of the quark-loop are not suppressed by $1/N_f$.  Moreover, the flavour-colour ratio is certainly of crucial importance for e.g.\ the question of the existence and size of a quarkyonic phase. Finally, we consider the following generic rescaling of  $\rho$, $V_k(\rho)$ and $h_k(\rho)$, 
\begin{align} \nonumber
&\, \hspace{6mm} \rho \to \frac{N_f^2-1}{N_\pi}\,\rho \,,\\[1ex] \nonumber
& V_k(\rho) \to \frac{N_f^2-1}{N_\pi}\,V_k(\rho)\,, \\[1ex] 
& h_k(\rho) \to \sqrt{\frac{N_\pi}{N_f^2-1}}\,h_k(\rho) \,.
\label{eq:rescalePi} \end{align}
\end{subequations}
The factor $N_\pi$ in \labelcref{eq:rescalePi} is introduced to simulate the flows of a quark-meson model with $N_\pi$ pions instead of one sigma meson and three pions. Both cases are relevant for the physics of two-flavour QCD or the two-flavour QM-model. 

In the chirally symmetric phase for large temperatures and cutoff scales, the pions and the sigma are degenerate on-shell at $\rho=0$. The second derivative term vanishes $\left. 2 \rho V''(\rho)\right|_{\rho=0}=0$, and the on-shell $\sigma$-propagator agrees with the pion one, and the (on-shell) flow equation resembles that with  four pions. 

In turn, in the broken phase, the $\sigma$-mode develops a mass and quickly decouples from the dynamics of the system. Then, the (on-shell) dynamics of the theory is driven by the three (massless or light) pions. From previous fRG investigations of the quark-meson model as well as QCD we know that the mesonic dynamics in the broken phase is of sub-dominant importance for not too large chemical potential. This suggests that the $N=4$ case should mimic the two-flavour case best. A full discussion 
of the comparison is provided in \Cref{sec:Asymptotic} and \Cref{sec:ChiralPhase}. 

With the limit $N_f \to \infty$ and \labelcref{eq:rescale} we derive the flow equations for  large-$N_f$ 
Yukawa coupling, $h_k^{\mathrm{lN}}(\rho)$, and effective potential, $V_k^{\mathrm{lN}}(\rho)$, 
\begin{align} \label{eq:udrho}\nonumber 
\partial_t V_k^{\mathrm{lN}}(\rho)= &\
\frac{k^5}{12 \pi^2} \Bigl\{ \frac{N_\pi}{\epsilon_k^\pi} (1 + 2n_B(\epsilon_k^\pi)) \\[1ex] 
&
- \frac{ 4 \times 2 \times 3}{\epsilon_k^q}(1  -n_f(\epsilon_k^q + \mu)-n_f(\epsilon_k^q - \mu)) \Bigr\}
\,,  	
\end{align}
and 
\begin{align} \label{eq:hdrho} 
\nonumber
\partial_t h_k^{\mathrm{lN}}(\rho)=&\,
-4 N_\pi v_3 (h^{\mathrm{lN}}_k(\rho))^3 L^{(4)}_{(1,1)}(m^2_{q,k},m^2_{\pi,k};T,\mu) \\[1ex]
&-2 N_\pi v_3 k^2 (h^{\mathrm{lN}}_k(\rho))' l_1^{(B,4)}(m^2_{\pi,k};T)
\,.
\end{align} 
This concludes our derivation of the set of flow equations solved in the present work: the system of flows at finite $N$ is given by \labelcref{eq:udrhoqm} and \labelcref{eq:hdrhoqm}, those in the large $N$-limit are given in \labelcref{eq:udrho} and \labelcref{eq:hdrho}. Numerical results for both systems will be presented in \Cref{sec:result}, the discontinuous Galerkin setup, with which the numerical results are achieved, are discussed in the next section.

\section{Discontinuous Galerkin Methods in the FRG}\label{sec:galerkin}

Most of the flow equations introduced in the previous \Cref{sec:qm} can only be solved numerically. In the present work we use Discontinuous Galerkin methods (DG-methods),  which have been introduced to the fRG in \cite{Grossi:2019urj} by the example of the large-$N$ limit in an $O(N)$-model. In contrast to the set of flow equations discussed in the present work for the QM-model, the flow equation for the effective potential in the large $N$-limit of the $O(N)$-model is given by an hyperbolic equation of first order that can be written in a conservative form.
This type of equation is well studied and understood. Many different numerical schemes were developed to obtain a stable solution in a weak sense see e.g. \cite{2013rehy.book.....R}.

In the present case, the flow equations \labelcref{eq:udrho} and \labelcref{eq:hdrho} in the large-$N$ limit are not conservative anymore. Indeed, for a constant Yukawa coupling the QM-model can be understood as a driven $O(N)$-model, where the driving force is provided by the quark loop. If this approximation to the effective action is upgraded to one with a cutoff-dependent quark two-point function, there are back-coupling effective from the meson loop into the quark-loop, and the driving-force is not (fully) independent anymore.  

In any case the system of differential equations ceases to be conservative. For the non-conservative hyperbolic problem, like the large $N_f$ equation with running Yukawa coupling, the concept of a weak solution was introduced relatively recent in \cite{pares2006numerical,castro2006high} and applied in a context of Finite Volume and Discontinuous Galerkin schemes \cite{hou1994nonconservative,CASTRO2016347, dumbser2010force, dumbser2009ader, castro2010some, castro2008many, Dumbser2010, cf7315f2fac84dc9beed7e7b65f54ec3} for multiple physical systems. Hyperbolic equations in non-conservative form occur rather frequently in modeling physical system, as example viscous relativistic hydrodynamic equations are of this type \cite{Bozek:2011ua,Schenke:2010nt,Karpenko:2013wva,Shen:2014vra,DelZanna:2013eua,Noronha-Hostler:2013gga,Floerchinger:2017cii, Floerchinger:2018pje} and recently also a formulation of general relativity has been solved in this formulation \cite{PhysRevD.97.084053} highlighting more advanced  stability properties. Moreover, in case of finite $N_f$ it also contains a diffusion term that originates in the $\sigma$-loop. Thus, on the technical level, the present work represents a non-trivial extension of \cite{Grossi:2019urj}. The different extensions are discussed in \Cref{sec:noncon} (non-conservative systems) and \Cref{sec:FiniteN} (diffusion terms). With respect to these extensions the present work should be considered as a first step to the full implementation of DG-methods for non-conservative systems, more details and further extensions will be considered elsewhere. For a more detailed introduction to DG-methods see also  \cite{Hesthaven:2007:NDG:1557392}.

In the context of the FRG, further work concerning the inclusion of higher order derivatives and non-conservative formulations has been achieved in~\cite{AdrianMartinInPrep:2021zzz}, which will be published soon.

Pseudo-spectral methods are an integral part of DG-methods. They are applicable to FRG equations in the absence of shocks, and have been used successfully in e.g.~\cite{Knorr:2016sfs,Borchardt:2016pif,Knorr:2017yze,Dabelow:2019sty,Knorr:2020rpm}.

\subsection{Non-conservative flux equations}\label{sec:noncon}

In this section the extension of DG-methods for the fRG to non-conservative flow equations is set up. To this end we consider a system of differential equations of the form, 
\begin{align}
\partial_t u_i+ \partial_\rho f_i(\bm u,\rho,t)+  a_{ij}(\bm u, \rho, t)\partial_\rho u_j = s_i(\bm u, \rho, t) \,,
\label{eq:pdeform}
\end{align}
where $\bm u=(u_1,u_2)^T$ and $i,j \in \{1,2\}$. The $s_i$ are source terms and $f_i$ conservative fluxes. In \labelcref{eq:pdeform} we also allow for non-conservative terms $a_i$. In the full quark-meson model the flux is additionally separated into a convective and a diffusive contribution depending also on $\partial_\rho u_i$. We note that the splitting into conservative and non-conservative terms is not unique in these equations. 
More details on the numerical treatment can be found in \Cref{app:numerics}, whereas an evaluation of the convergence properties of the scheme is performed in \Cref{app:conv}.
In the following \labelcref{eq:udrhoqm}, \labelcref{eq:udrho} and \labelcref{eq:hdrho} are reformulated to fit \labelcref{eq:pdeform}. 

\subsubsection{Flow of the effective potential}\label{sec:derive}

\Cref{eq:udrhoqm} and \labelcref{eq:udrho} are rearranged to ensure that they have the conservative form required by \labelcref{eq:pdeform} such that DG methods are applicable in a purely conservative formulation.
Similarly to \cite{Grossi:2019urj} we observe a non-linear dependence of the potential $V_k(\rho)$ on its derivative with respect to the field expectation value $\partial_\rho V_k(\rho)$ and in case of \labelcref{eq:udrhoqm} also on its second derivative. The dependency on the first derivative can be eliminated by introducing it as a new variable, which coincides with the pion mass squared:
\begin{align}\label{eq:newu}
u_k(\rho) = \partial_\rho V_k(\rho) = m^2_{\pi,k} \,.
\end{align}
Since \labelcref{eq:udrho} is not dependent on itself we can simply take a $\rho$ derivative, which turns $u_k(\rho)$ into a conserved quantity which is fit for DG schemes. 
This procedure is also applied to the QM-model. In this case we need an additional expression for the second derivative of the potential in the sigma mass in \labelcref{eq:mpsi}. We obtain this expression by taking another $\rho$ derivative of the polynomial basis functions $\phi_n$ introduced in \Cref{app:numerics}:
\begin{align}\label{eq:polyn}
\partial_\rho^2 V(t,\rho)=\partial_\rho u(t,\rho) = \displaystyle \sum_{n=1}^{N+1} \hat{u}_n(t)\partial_\rho (\phi_n(\rho)) \,.
\end{align}

\subsubsection{Flow of the field-dependent Yukawa coupling}
The flow of the Yukawa coupling at finite N is given by a highly non-linear equation of second order. Since it can not be made to fit the form given in \labelcref{eq:pdeform}, it is not solved within the introduced framework.
However, the expression simplifies significantly in the large-N limit and \labelcref{eq:hdrho} can be written to suit the formalism.
\Cref{eq:hdrho} is rewritten in terms of the quark mass squared $m_q^2(\rho)$, as we are primarily interested in physical observables. Thus, a new variable is introduced, 
\begin{align}\label{eq:newv}
w_k(\rho) = 2h_k(\rho)^2\rho  = m_{q,k}^2 \,.
\end{align}
We refer to \Cref{app:flow} for the explicit calculation. Introducing this new variable proves to be very helpful for the computation.
\Cref{app:non_cons_flux} explains how the ambiguity in splitting the conservative and non-conservative contributions to the flux are used to accommodate boundary conditions. For completeness the final form of the equations is stated,  
\begin{align}\nonumber
\partial_t u_k=&\, \partial_\rho f_{u}(u_k,w_k) \,,
\\[1ex] 
\partial_t w_k=&\, \partial_\rho (a(u_k) w_k) - (w_k \partial_u  a(u_k)) \partial_\rho u_k+ s(u_k,w_k) \,.
\label{eq:fin}
\end{align}
This version of the equation has the advantage that the non-conservative product is rather small in comparison to the conservative part.

\subsection{Finite N equations}\label{sec:FiniteN}

For finite N, the equation is parabolic. Apart from the convection term (Goldstones), there is also a diffusion term that arises form the $\sigma$-loop. The equations for finite $N$ are highly non-linear and of second order, therefore we refrain from considering the field dependency of the Yukawa coupling. Schematically the flow equation of the potential is written as 
\begin{align}
\partial_t u_k=\, \partial_\rho f_{u}(u_k,w_k , \partial_\rho u_k ) \,.
\label{eq:full} 
\end{align}
The weak formulation and the stability of this type of equation has not been fully understood until now. The presence of the diffusion modifies the numerical flux significantly. However, in the convection dominated regime, and in the absence of a discontinuity, it is possible to neglect this diffusion numerical fluxes and formulate the Discontinuous Galerkin method for this equation as follows, 
\begin{align} \nonumber
	\int_{D^k} \Big( (\partial_t u_{i,h}+  a_{i,h}\partial_\rho u_{i,h} + s_{i,h})q_n  + f_{i,h} \partial_x q_n \Big)d x \\[1ex]
	= - \int_{ \partial D^k}  q_n  \Big(f_i^* \hat{\mathbf{n}} + \mathbf{D}(u_{i,h}^+, u_{i,h}^-, \hat{\mathbf{n}}) 
	\Big) dx \,.
	\label{eq:weakFIniteN }
\end{align}
where $a_{i,h}$, $s_{i,h}$ and  $f_{i,h}$  are computed form the field $u_i$ and their local approximation of the derivative, no other numerical fluxes are introduced into the numerical scheme. The absence of numerical fluxes for the extra derivative present in the equation correspond to the assumption of continuity of this field and the DG scheme somehow reduces to a pseudo spectral method. This approximation is acceptable, whenever the flow is rather smooth and no shock or rarefaction wave are generated during the simulation. In turn, this scheme will fail in the vicinity of a first order phase transition. There we expect shock-formation and propagation in the flow equation. In conclusion, for the rest of the phase diagram the present approximation can be considered as a sufficiently accurate solution of the flow equation due to the local high order accuracy of the DG scheme.

\section{Results}\label{sec:result}

In this section we present and discuss our results for the phase structure of the QM-model in the different approximations. We chose our initial conditions such that the quark mass and the pion decay constant reproduce physical values in the vacuum, this is discussed in detail in \Cref{sec:InitialConditions}. Note however, that its not the main objective of the present work to produce quantitatively reliable results, the vacuum bench marking simply facilitates the understanding of our results. In the present work we rather focus on the qualitative behavior of the matter sector of QCD at large densities, quantitatively reliable results require full QCD flows and will be considered elsewhere. Such a set-up entails, that while we compute and present a phase structure at large densities, our present low energy effective theory gradually looses predictability for  larger $\mu_B/T$. Such an estimate in functional QCD leads to a predictability bound of $\mu_B/T \lesssim 4$, if the currently existing state of the art computations are combined, \cite{Fu:2019hdw, Gao:2020fbl, Gao:2020qsj} and estimates for missing channels and effects are considered as well, \cite{Fu:2019hdw, Braun:2019aow, Eichmann:2015kfa}. In the present class of low energy effective theories (QM, NJL-type, PQM, PNJL), a respective estimate leads to $\mu_B/T \lesssim 2$.

We first present results within an approximation where only the effective potential depends on the cutoff scale, the \textit{local potential approximation} (LPA), for both, the finite $N_f$ and the large-$N_f$ limit in \Cref{sec:reslN}. They serve as a benchmark for the more advanced approximations discussed in \Cref{sec:resy}. 
Additionally, the results in \Cref{sec:reslN} also serve as benchmark for results in the literature within the QM- and Polyakov loop-enhanced QM models, in particular at large density, where DG-methods or similar numerical approaches are mandatory for reliable results. 

In \Cref{sec:resy} we present results for the coupled system of effective potential $V_k(\rho)$ and Yukawa coupling $h_k(\rho)$ in the large-$N_f$ limit. This investigation allows us to solidify the results in \cite{Pawlowski:2014zaa} concerning the flattening of the quark mass $m_q(\phi)$. 

Lastly, the technical advances made here readily carry over to first principle QCD within the fRG, as discussed in the introduction, they are one of two missing ingredients for reliable predictions of the QCD phase structure for $\mu_B/T\gtrsim 4$.

\subsection{Initial Conditions}\label{sec:InitialConditions}
\begin{table}[t]
	\begin{center}
		\begin{tabular}{|c | c || c |}
			\hline  & & \\[-1ex]
			Observable	& Value [MeV]  & Parameter at $\Lambda_{UV} = 0.65$\,GeV \\[1ex]
			\hline \hline &  & \\[-1ex]
			$m_\sigma$& 317.1  & $\lambda_\Lambda$ = 71.6  \\[1ex]
			$m_q$& 310.8 & $h_{\phi,\Lambda}= $ 3.6   \\[1ex]
			\hline 
		\end{tabular}
	\end{center}
\caption{Low energy observables and related  EFT couplings at the initial cutoff scale $\Lambda_\textrm{UV}=0.65$\,GeV. The scales are fixed with the pion decay constant in the chiral limit $f_{\pi,\chi} = 88$\,MeV, that is $m_\sigma/f_{\pi,\chi}\approx 3.603$ and $m_q/f_{\pi,\chi} \approx 3.532$. In the present approximation we have $f_{\pi,\chi} = \sigma_0$, and in the model the  dimensionless ratios are simply $m_\sigma/\sigma_0$ and $m_q/\sigma_0$. In the chiral limit we also have $m_\pi=0$.\hspace*{\fill} }
\label{tab:ini}
\end{table}
We initiate the flow at a cutoff scale $k=\Lambda\approx 0.650$\,GeV with the classical action of the QM-model. Then, the parameter in the initial effective action $\Gamma_\Lambda$ is the $\phi^4$-coupling in the classical potential, 
\begin{align}
\label{eq:init}
	u_\Lambda (\rho) = \frac{\lambda_\Lambda}{2} \rho \hspace{0.6cm}
	w_\Lambda (\rho) = 2 h^2_\Lambda \rho 
\, .
\end{align}
as well as the Yukawa coupling $h_\Lambda$. For the sake of simplicity we use a initial meson quark mass, $m_{\phi}^2=0$.  
All our scales are measured in the pion decay constant in the chiral limit $f_{\pi,\chi}= 88$\,MeV. Within the present approximation of the QM-model we have $f_\pi\approx \sigma_{0}$, and hence we define $\sigma_{0}= 88$\,MeV. Then, the two model parameters $\lambda_\Lambda, h_\Lambda$ are fixed such that they lead to a 'physical' constituent quark mass $1/\sqrt{2} h \sigma_0$, and a 'physical' mass of the sigma resonance, $m_\sigma$. The parameters for the couplings of the effective theory and their relation to physical observables are summarised in \Cref{tab:ini}. The dimensionless ratios in the models at $k=0$ are given by  
\begin{align} 
\frac{m_\sigma}{\sigma_{0}} \approx 3.605\,,\qquad  \frac{m_q}{\sigma_{0} }\approx 3.532\,, 
\end{align}%
and follow with the initial parameters in \Cref{tab:ini}.

\subsection{Results for the effective potential with constant Yukawa coupling}\label{sec:reslN}

In this section we compare the numerical results of the physical case with  $N_f=2$, and in the large-$N_f$ limit with three and four degrees of freedom. This is done in LPA, where we solve the flow equation for the effective potential, \labelcref{eq:udrho}. We first discuss the asymptotic regimes: vacuum, large temperatures, and large chemical potential, \Cref{sec:Asymptotic}. Then we show that the chiral phase transition, or rather its non-universal properties, agree quantitatively for all models, \Cref{sec:ChiralPhase}. The shock-development at large chemical potential is discussed in \Cref{sec:Shock}. Finally, we compare the phase structure for all three cases in \Cref{sec:PhaseLPA}.

\begin{figure}[t!]
	\centering
	\includegraphics[width=\linewidth]{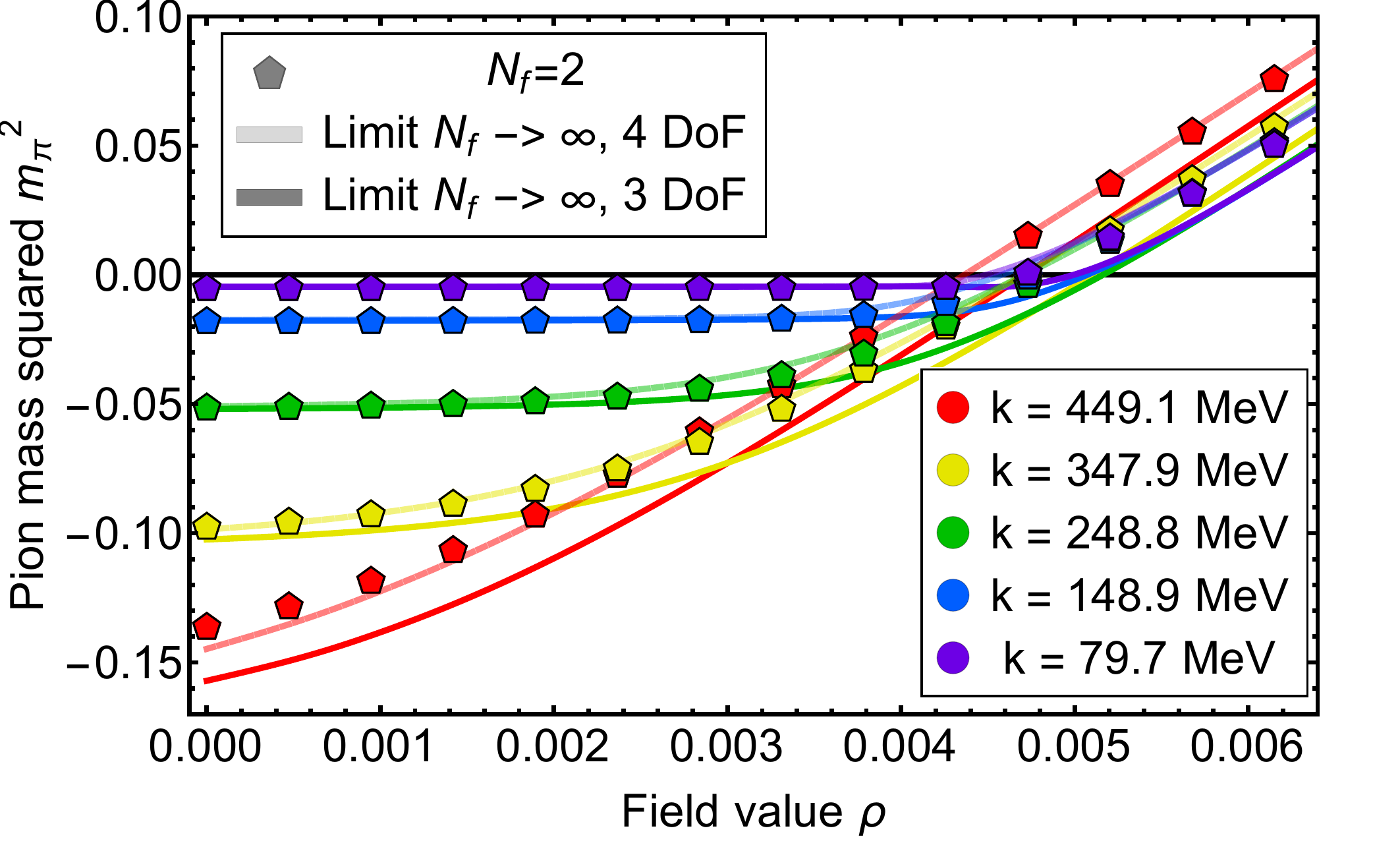}
	\caption{RG-scale evolution of the field-dependent pion mass $m_\pi^2(\rho)$ in LPA (constant Yukawa coupling) in the vacuum. We compare results for $N_f=2$, and the large-$N_f$limit with 3 and 4 degrees of freedom, see \Cref{sec:largeN}. \hspace*{\fill} }
	\label{fig:00}
\end{figure}
%

\subsubsection{Asymptotic regimes} \label{sec:Asymptotic}

We begin with an evaluation of the asymptotic regimes: the vacuum with $\mu_q,T =0$, large temperatures with $\mu_q=0$, and large chemical potentials with $T=0$. For these cases we show the field-dependence of the pion mass $m_{\pi,k}(\rho) =\partial_\rho V_k(\rho)$, see \labelcref{eq:newu} for different cutoff scales. This resolves the effective potential,  obtained from an integration over $\rho$, in terms of a physical observable. 

For the numerical solution of \labelcref{eq:udrho} we use an interval of $\rho \in [0, 0.02]$\,GeV, which is expanded in $K=80$ cells with polynomials of order $N_p=2$. The length of the interval is chosen such that it includes all relevant phenomena: the flux at the outer boundary is very small. This ensures the numerical convergence for the entire phase diagram. 

The benchmark case is the vacuum, where the present models are anchored, see \Cref{sec:InitialConditions}. The field-dependence of the pion mass is shown in \Cref{fig:00}. For the initial cutoff $k=\Lambda$, the pion mass is simply a 
straight line, $m_{\pi,\Lambda}(\rho)= \lambda_\Lambda\rho$, where the slope is the initial mesonic self-coupling, $\lambda_\Lambda$. With decreasing cutoff scale, the pion mass develops a flat regime, which is related to the emergence of non-trivial minima $\rho_0=\sigma_0/2$ in the potential, indicating chiral symmetry breaking.  

We also conclude, from the comparison of the pion masses in the different models, that the cutoff-dependence of the physical two-flavour case is best mimicked by the large-$N_f$ limit with four degrees of freedom: most of the dynamics takes place in the symmetric regime or close to it. Technically, this regime is described with $1+m_\sigma^2(\rho_0)/k^2\approx 1+m_\pi^2(\rho_0)/k^2$, owing to the fact that the total mass of the respective modes is $k^2+m^2_{\pi/\sigma}$. 
\begin{figure}[t!]
	\centering
	\includegraphics[width=\linewidth]{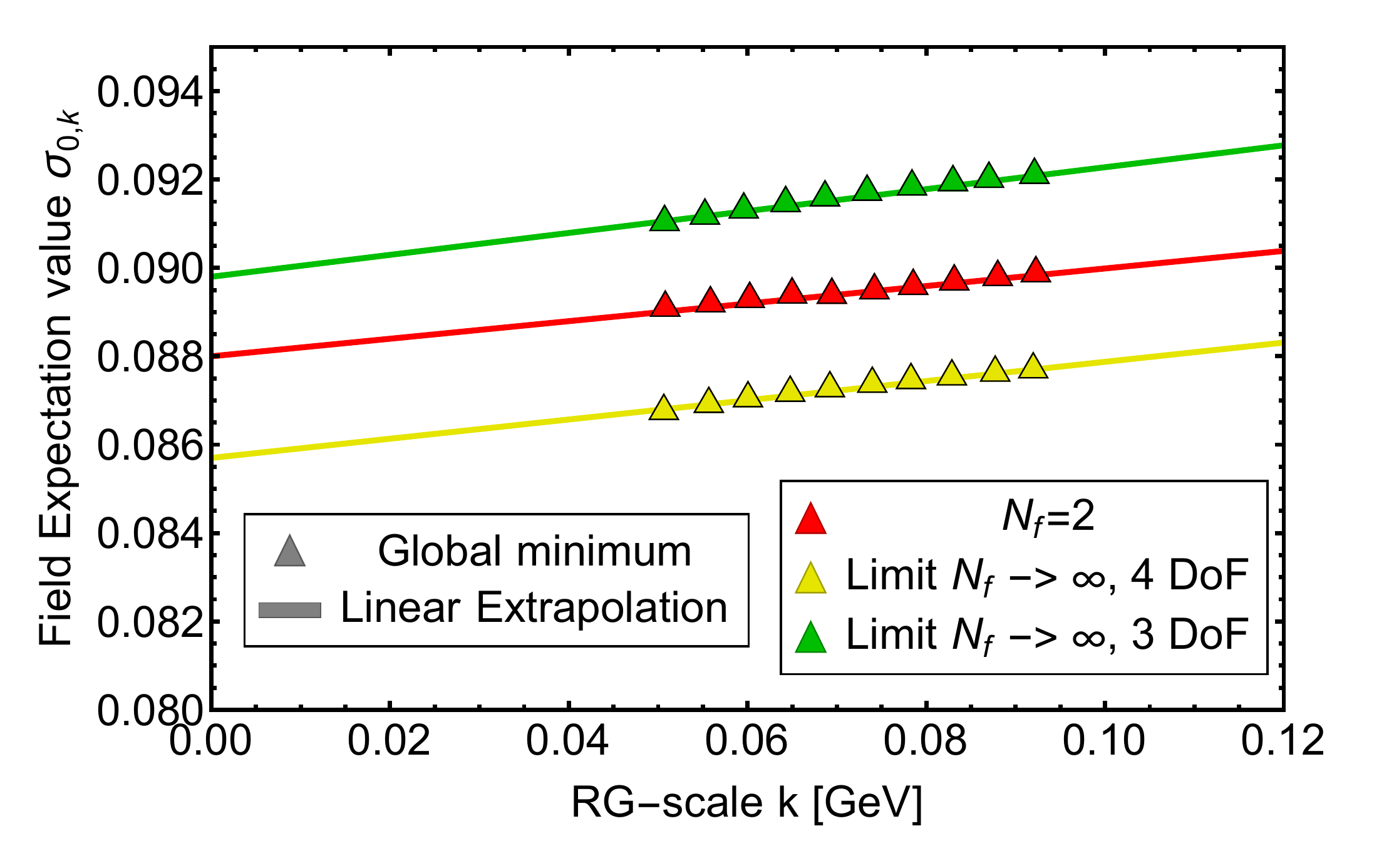}
	\caption{Linear extrapolation of the expectation value $\sigma_0$. The corresponding fit parameters to \labelcref{eq:fitpd} can be found in \Cref{tab:rho0}.\hspace*{\fill} }
	\label{fig:Linfit}
\end{figure}

Finally, we determine the relative values of the pion decay constants in the chiral limit, $f_{\pi,\chi}=\sigma_0$, in the different models. All scales are measured in the pion decay constant $f_\pi=88$ in the two-flavour case. The expectation value $\sigma_0$ or $\rho_0=\sigma_0^2/2$ follows from \labelcref{eq:mphi} as the maximal field value with $m_\pi(\rho_0)=0$. However, since we stop the numerics at a small but finite cutoff value, $k_\textrm{min}=50$\,MeV, we extrapolate the expectation value $\sigma_0(k_\textrm{min}) $ to $\sigma_0(0)$ within a linear fit: we use data from 10 equally spaced RG-scales from $k=90$\,MeV to $k=50$\,MeV, and fit 
\begin{align}
\sigma_{0,k}= \sigma_{0} + const \ k \,.
\label{eq:fitpd}
\end{align}
The self-consistency of this linear fit is checked by the perfect agreement of the linear fit with the data, see \Cref{fig:Linfit}. The respective values for $\sigma_0$ are given in \Cref{tab:rho0}. The mass $m_\sigma$ of the scalar mode is extrapolated to $k=0$ from the same data. Note however, that once the kink enters the cell in which $\sigma_0$ is located, the precise determination of the derivative is difficult. Therefore, the flattening of the potential most likely causes an underestimation of $m_\sigma$.

\begin{figure*}[t!]
	\centering
	\begin{minipage}[b]{0.49\linewidth}
		\includegraphics[width=\linewidth]{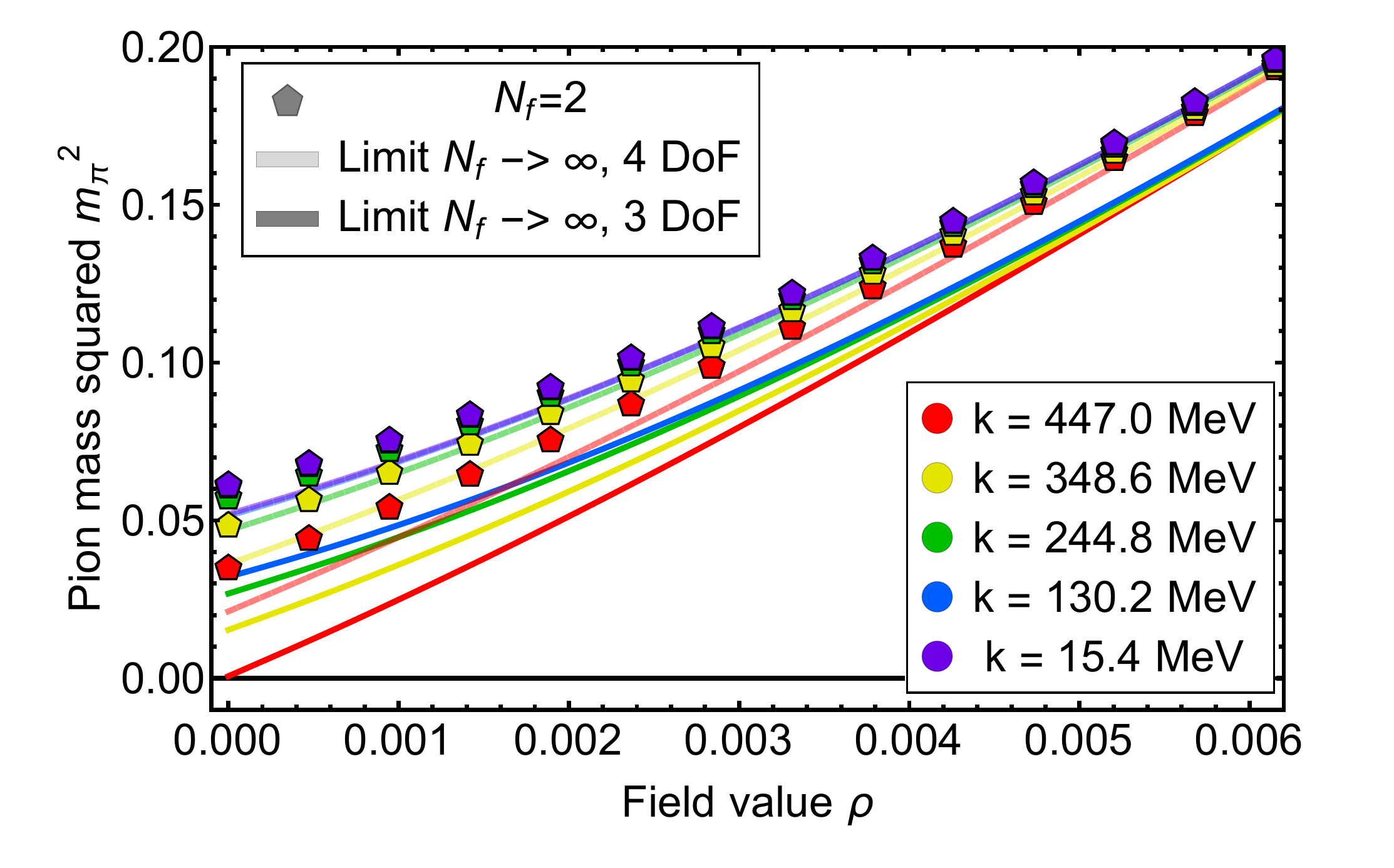}
		\subcaption{Pion mass at zero density and high temperature ($T=280$ MeV). The flow is washed out by the temperature fluctuations.\hspace*{\fill} }
		\label{fig:30}
	\end{minipage}%
	\hspace{2mm}%
	\begin{minipage}[b]{0.49\linewidth}
		\includegraphics[width=\linewidth]{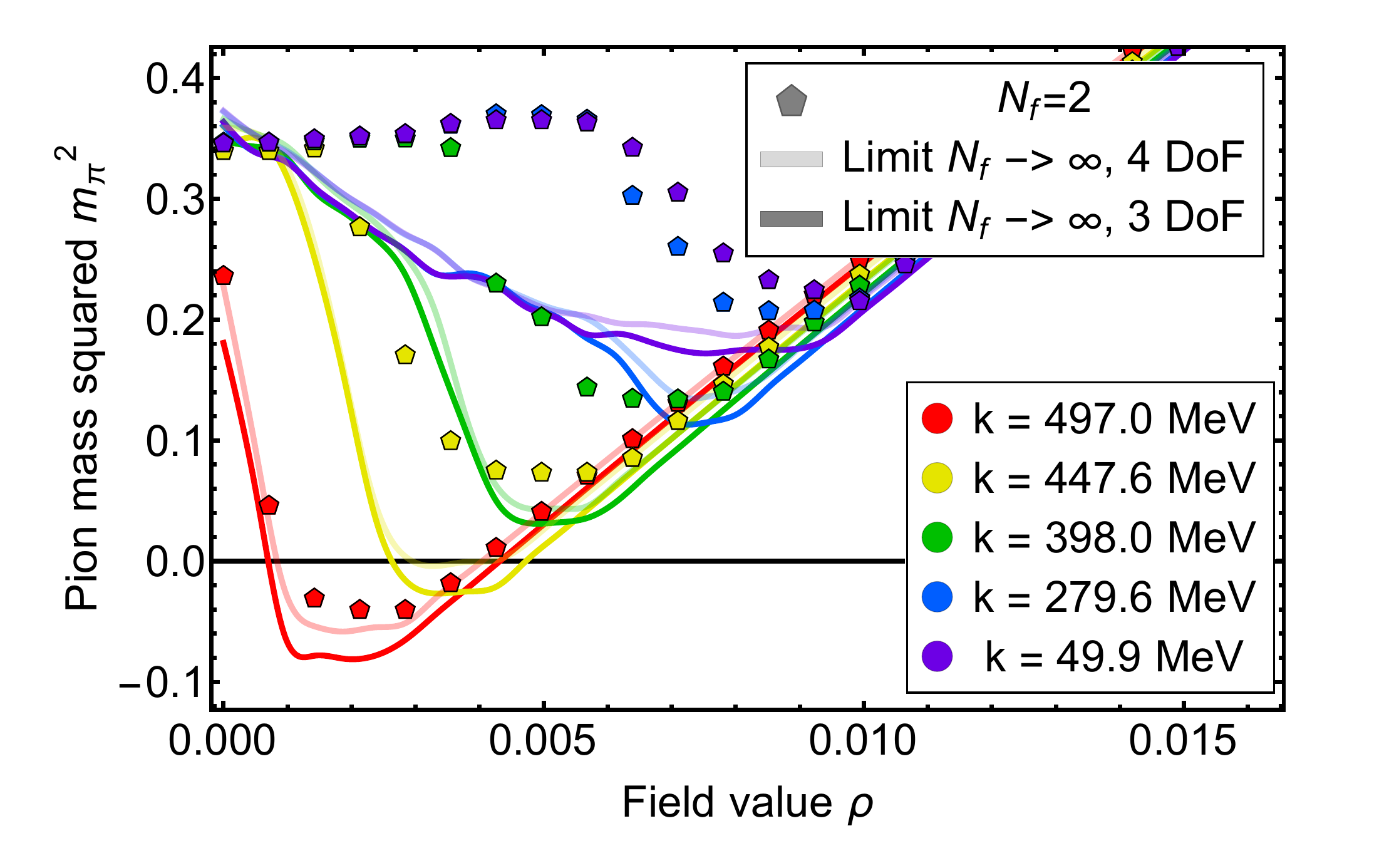}
		\subcaption{Pion mass at zero temperature and high density ($\mu = 500$ MeV). The density onset introduces a sharp edge to the flow.\hspace*{\fill} } 
		\label{fig:05}
	\end{minipage}	
	\caption{RG-scale evolution of the field dependent pion mass $m_\pi^2$ in the symmetric phase. In this figure we compare results for the finite $N_f$ case and the large $N_f$ limit with 3 and 4 degrees of freedom respectively. The Yukawa coupling is kept at a constant value.\hspace*{\fill} }	
\end{figure*}

\begin{table}[t]
	\vspace{.3cm}
	\centering
	\begin{tabular}{|l||c|c|c|} 
		\hline
		&  & &\\[-1ex]
		& $\sigma_0$ [MeV] & const & $m_\sigma$ [Mev] \\[1ex]
		\hline \hline\hspace{2.4cm} & \hspace{1.6cm} &\hspace{1.6cm} &\hspace{1.6cm}\\[-1ex]
		$N_f \to \infty$, 3 DoF  & 89.8(17) &0.0755(32) &335(15)  \\[1ex] 
		$N_f \to \infty$, 4 DoF & 85.7(18) &0.0752(35)& 311(10) \\[1ex] 
		$N_f = 2$ & 88.0(20)& 0.0360(26)& 317(12)  \\[1ex] 
		\hline 
	\end{tabular} 
	\caption{Expectation value $\sigma_0$ for the three models obtained by a fit of \labelcref{eq:fitpd} to the zero point at 5 equally spaced RG-scales $k=90$ MeV to $k=50$ MeV. The error is computed from the grid resolution and the error to the fit parameters. The mass $m_\sigma$ is extrapolated from the same data points as the fit. The error of $m_\sigma$ is obtained analogously to the one of $\sigma_{0}$, it might be underestimated due to the kink developing at $\sigma_{0}$.\hspace*{\fill} }\label{tab:rho0} 
\end{table}
For large temperatures we safely stay in the symmetric regime and the mesons simply acquire a thermal Debye mass. This is seen in \Cref{fig:30}. In the symmetric regime we have four mesonic degrees of freedom in the two-flavour case. Consistent with our expectations, that the cutoff-dependence of the pion mass in the model with four degrees of freedom in the large-$N_f$ limit has the best agreement with the two-flavour case. 

We close with a discussion of the large chemical potential asymptotics. The respective pion mass (squared) is depicted in \Cref{fig:05}. The sudden increase of the pion mass for $k\lesssim k_\mathrm{on}$ with $k_\mathrm{on} \approx \mu$ in \Cref{fig:05} is related to the Silver Blaze property, \cite{Cohen:2003kd}, for the discussion in the fRG-approach see \cite{Khan:2015puu, Fu:2015amv, Fu:2016tey}. This property entails that correlation functions below the density onset are simply functions of $p_0\mp i \mu_q$ for quark and anti-quark frequencies respectively. Accordingly, observables do not depend on the chemical potential for $\mu_q<\mu_{q,\textrm{on}}$, where $\mu_{q,\textrm{on}}$ is the onset chemical potential. For $\mu_q>\mu_{q,\textrm{on}}$, the medium leads to deformations of the quark-meson scattering processes, comprised in medium meson-dispersions. In the presence of thermal fluctuations this onset is washed out with increasing temperature.  

Note also, that the onset cutoff scale depends on the field value, as in the present approximation the onset chemical potential is given by $\mu_{q,\textrm{on}}^2 = 1+ m_{q,k}^2$. 

In summary, the results in the asymptotic regimes show the expected physics phenomena. Moreover, the comparison of the large-$N_f$ models with the two-flavour case quantifies the similarities between the large-$N_f$ limit models and the physical two-flavour model.

\subsubsection{Chiral phase transition at vanishing density}\label{sec:ChiralPhase}

\begin{figure*}[ht!]
	\begin{minipage}[b]{0.49\linewidth}
		\includegraphics[width=\linewidth]{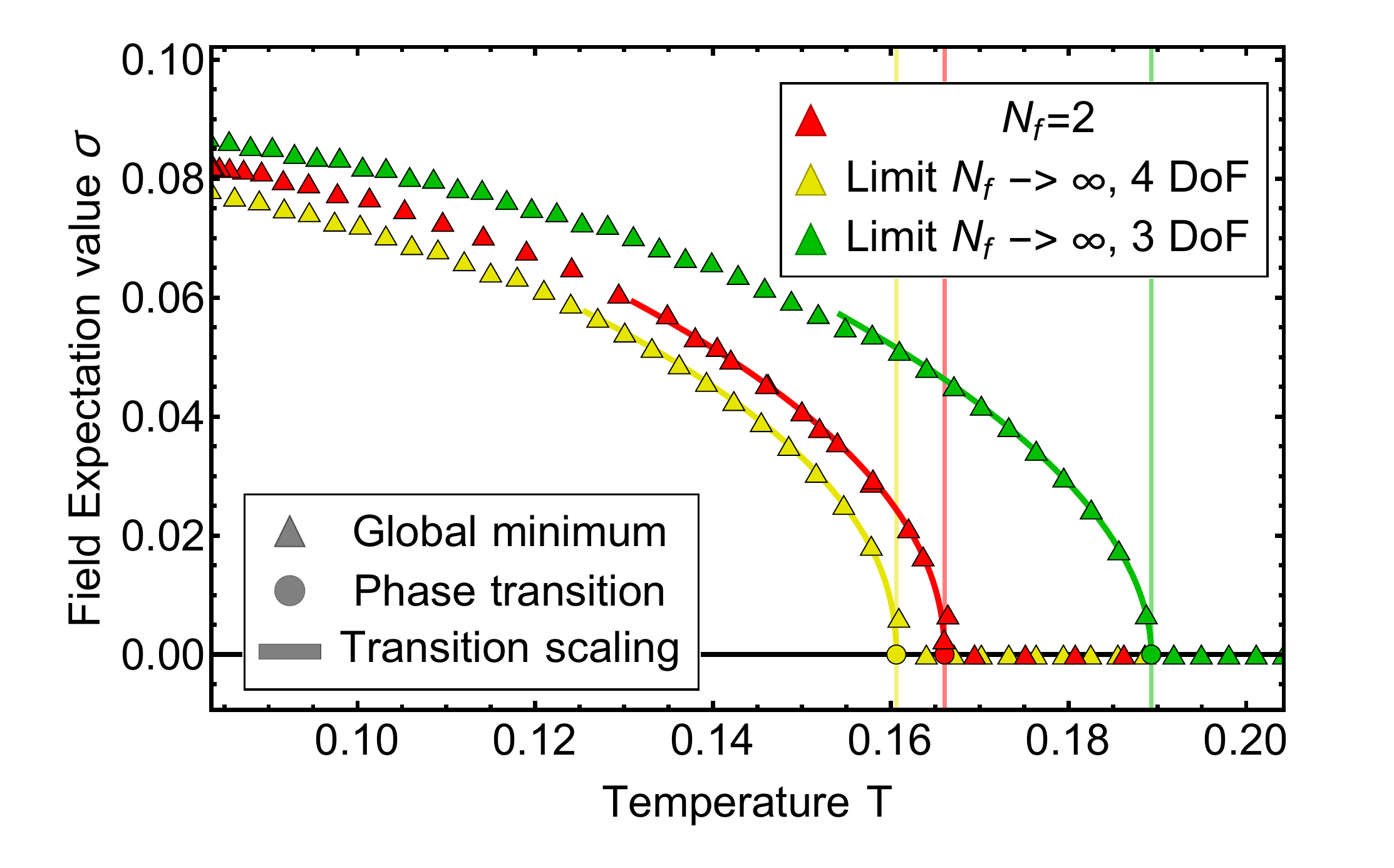}
		\subcaption{Absolute values of the field expectation value $\sigma$. Units are given in GeV.\hspace*{\fill} }
		\label{fig:firstorder}
	\end{minipage}\hspace{2mm}
	\begin{minipage}[b]{0.49\linewidth}
		\includegraphics[width=\linewidth]{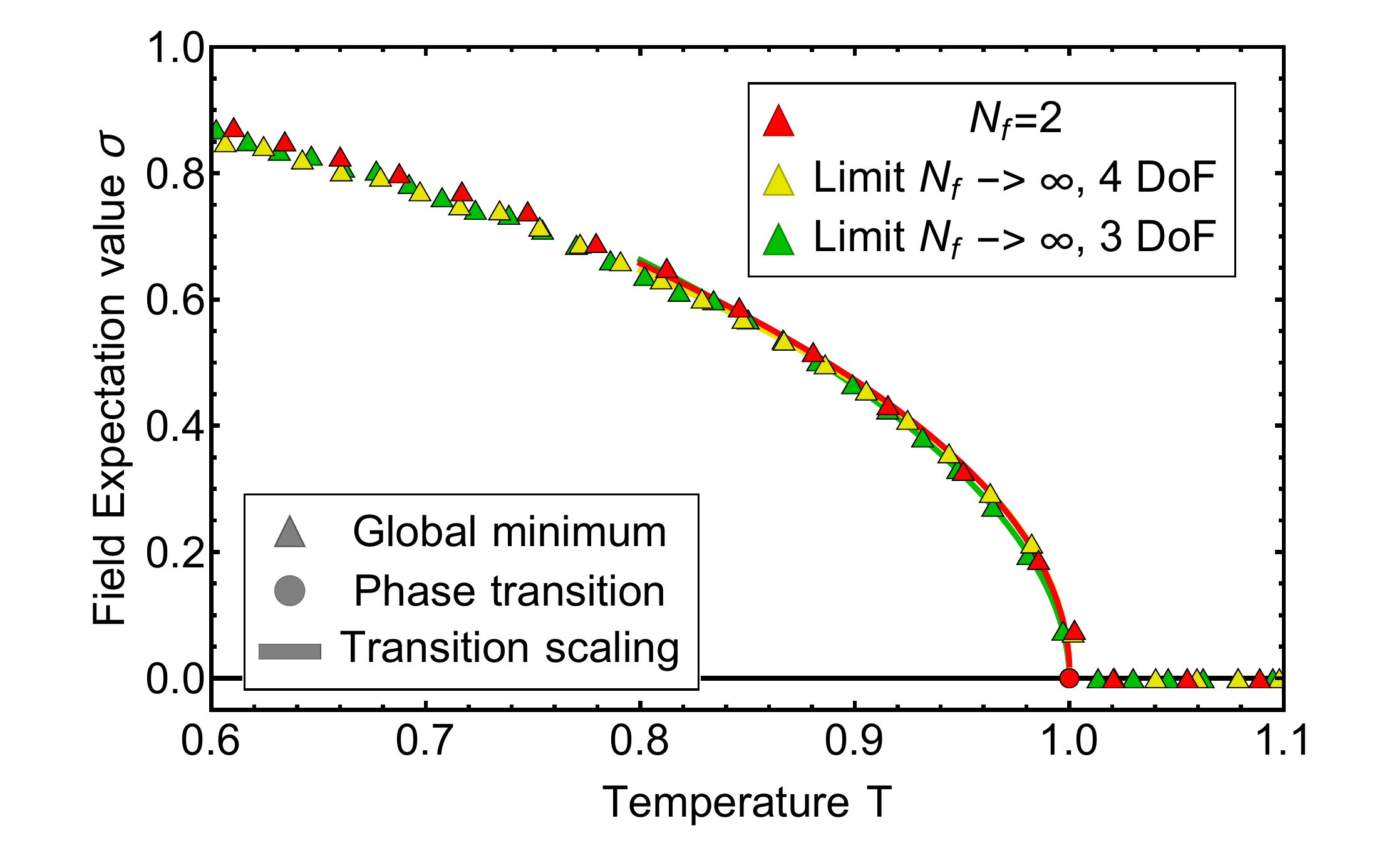}
		\subcaption{Axes rescaled by the respective expectation values and critical temperatures.\hspace*{\fill} }
		\label{fig:scaling}
	\end{minipage}	
	\caption{Temperature dependence of the mesonic field expectation value $\sigma_0$. The figure shows a second order phase transition at fixed chemical potential $\mu = 0$\,GeV. We compare the case for finite $N_f$ and the large-N limit with 3 and 4 degrees of freedom with a constant Yukawa coupling. Fitting \labelcref{eq:fit1} to the data gives values for the critical temperatures and exponents of the transition, with the parameters given in \ref{tab:fit1}. Deviations to the fit values are due to the low resolution in the extrapolation of the zero point on the numerical grid.\hspace*{\fill} }
\end{figure*}

These similarities are furthered by a study of the chiral phase transition at vanishing density, $\mu_q=0$. In particular we present a detailed comparison of the temperature-dependence of the chiral order parameter $\sigma_0$ in the $N_f=2$-flavour case with the large $N_f$ limits with 3 and 4 degrees of freedom. The numerical results for $\sigma_0(T)$ are displayed in \Cref{fig:firstorder} and \Cref{fig:scaling}. 

In these computations we use a grid $\rho=[0,0.2]$\,GeV and expand in $K=100$ elements with a local approximation order of $N_p=2$ and the $\sigma$ is obtained by a linear extrapolation as described in \Cref{sec:reslN}. The data in proximity of the transition point are compatible with the scaling law, 
\begin{align}
\sigma(T) = 
\begin{cases}
c_\textrm{cr} |T - T_{\mathrm{crit}}|^\beta, \qquad \qquad & T \leq T_{\mathrm{crit}} \\[1ex]
0, & T \ge T_{\mathrm{crit}}
\end{cases} \,, 
\label{eq:fit1}
\end{align}
where for $N_F\to\infty$ we have $\beta =1/2$, the mean field critical exponent. In turn, for the present LPA-study of the Yuakwa model with the O(4)-universality class we have used the three-dimensional spatial flat or Litim regulator, \cite{Litim:2002cf}, for both quarks and mesons.  See also \Cref{app:thfkt}, \labelcref{eq:regphi}, \labelcref{eq:regq}. This leads us to $\beta\approx 0.40$, see~\cite{Bohr:2000gp} with \cite{Litim:2002xm, Litim:2002hj} and in particular the recent work in the QM-model, \cite{Chen:2021iuo}. Note, that more advanced approximations of the fRG provide $\beta \approx 0.39$ consistent with conformal bootstrap and Monte-Carlo results. For a recent compilation see \cite{DePolsi:2020pjk}, for the QM-model see \cite{Chen:2021iuo} that also includes an investigation of the $Z_2$-universality class. 

The respective scaling regimes are already very small in the $O(4)$-model and even shrink in the presence of the (driving) fermion loop, see the discussion in \cite{Chen:2021iuo}. While possible, we do not aim at a precision estimate of critical exponents here, as we focus on the location of the phase boundaries. Accordingly, we have simply checked the consistency of the scaling law \labelcref{eq:fit1} with $\beta \approx 0.4$ ($N_f=2$) and $\beta =1/2$ ($N_f\to \infty$) for small reduced temperatures $1-T/T_c \to 0_-$. This also allows us to determine the respective scaling regimes. Consistent with the observation above that they should be even smaller as the already small scaling regime in $O(N)$-models we find scaling for 
\begin{align}\label{eq:scalreg}
0> 1-T/T_c\gtrsim 10^{-2}\,. 
\end{align} 
Moreover, a scaling fit with \labelcref{eq:fit1} in the regime \labelcref{eq:scalreg} allows us to determine $T_c$ as well as the prefactor $c_\textrm{cr}$. 
\begin{table}[b] 
	\centering
	\begin{tabular}{|c||c| p{0.25\linewidth}||c|p{0.1\linewidth}|}
		\hline 
		 & & \\[-1ex]
		Model &  $c_\textrm{cr}$ & $T_{\mathrm{crit}}$\\[1.5ex] 
		\hline \hline
		 & & \\[-1ex]
		 $N_f \to \infty$: 3 DoF & 0.2985(24) & 0.18929(11) \\[1ex]
		 $N_f \to \infty$: 4 DoF & 0.3010(33) & 0.16139(10) \\[1ex]
		 Finite $N_f$ & 0.2126(61) & 0.16618(19) \\[1ex]
		\hline
	\end{tabular}
	\caption{Parameters obtained from a $\chi^2$-fit of \labelcref{eq:fit1} to the mean field expectation values in \Cref{fig:firstorder} which are underlined by the transition scaling. The error in the data is expected to be higher, since the numerical precision is limited by the grid resolution. An exact reconstruction of the zero-crossing is not possible.\hspace*{\fill} }
	\label{tab:fit1}
\end{table} 

We see from \Cref{tab:fit1}, that the two-flavour critical temperature agrees well with the large-$N_f$ limit with four degrees of freedom. This is expected from the theoretical analysis and our results on the asymptotics in \Cref{sec:Asymptotic}. This good agreement extends to the full temperature dependence, as can be seen from \Cref{fig:firstorder}. In turn, the order parameter from the large-$N_f$ limit with three degrees of freedom seemingly shows a slightly different behaviour. 

However, the two large-$N_f$ models are obtained by a simple rescaling of the fields and hence are identical to each other. They can be mapped onto each other by the relative rescaling. Put differently, the temperature dependence of the order parameters should agree if plotted in dimensionless units, $\sigma(T)/\sigma(0)$ and $T/T_c$. This comparison is shown in \Cref{fig:scaling}: as expected, the temperature-dependence of the order parameter of the large $N_f$ models agree. More importantly, also the two-flavour case agrees quantitatively, though with small deviations. Trivially, the non-trivial critical scaling of the two-flavour case does not agree with the trivial mean-field scaling for $N_f\to\infty$, but the scaling regimes are very small, see \labelcref{eq:scalreg}. 

In summary, the thermal properties of the models in the large $N_f$-limit and the physics case $N_f=2$ agree very impressively. 
\subsubsection{Shock Development and First Order Phase Transition at High Densities }\label{sec:Shock}

In this section we discuss the shock development and propagation at intermediate densities and very low temperatures. This is also used to discuss the first order regime. 

In these computations we use a grid $\rho=[0,0.2]$ GeV and expand in $K=200$ elements with a local approximation order of $N_p=2$. This finer grid is required for the shock resolution at low temperatures and chemical potentials close to the onset chemical potential. Indeed, the full resolution of some of the features in this regime (e.g.\ the precise location of the transition line in the absence of shocks) requires an even higher resolution. While technically possible, we have refrained from doing so, as the related aspects have been not in the main focus of the present work.  

We first note, that the running of the pion mass stops quickly below the onset RG-time $t_\mathrm{on} = \ln \frac{\Lambda}{\mu}$: the RG-flow  is proportional to the Fermi-distribution. Hence it stops at $k_\textrm{on}$ at $T=0$, for finite $T$ the Fermi-distribution is softer but for small temperatures there is still a strong exponential suppression for $k\leq k_\textrm{on}$. This suppression leads to two competing effects at finite densities:   
\begin{itemize}
	\item The onset amplitude, and thus $m_\pi^2 (\rho_0)$ in the symmetric phase, are linked to the suppression of the quark-contribution. This contribution dominates initially, but due to the constant Yukawa coupling it is quickly suppressed with $k^5$.
	
	\item For field values with positive meson masses $m^2_{\pi, k}(\rho)$ the meson loop in the flow is suppressed with $k^5$. In turn, for negative meson masses the meson loop is suppressed with $k^4$. Note also, that the flow increases with decreasing values of $m^2_{\pi, k}$, which is closely linked to the restoration of convexity. The mesonic flow contribution is reminiscent of the spreading of waves in hydrodynamics, where its value corresponds to the wave velocity: if we consider the solution $m_\pi^2(\rho)$ as a wave packet, it flows with the RG-time in the direction of smaller field values with a $\rho$-dependent propagation velocity. The velocity of the solution and its effect on convexity is inspected more closely in \Cref{app:convex}.
\end{itemize} 
The interplay of both effects leads to the creation of shocks and a first order phase transition at low temperatures. The increased propagation speed of negative modes is blocked by the slowed propagation of positive modes. The shock travels towards smaller field values during the RG-time evolution, but eventually freezes when the shock amplitude is too high. An illustration of this process can be found in \Cref{app:shock}. 

\begin{figure}[t]
	\includegraphics[width=\linewidth]{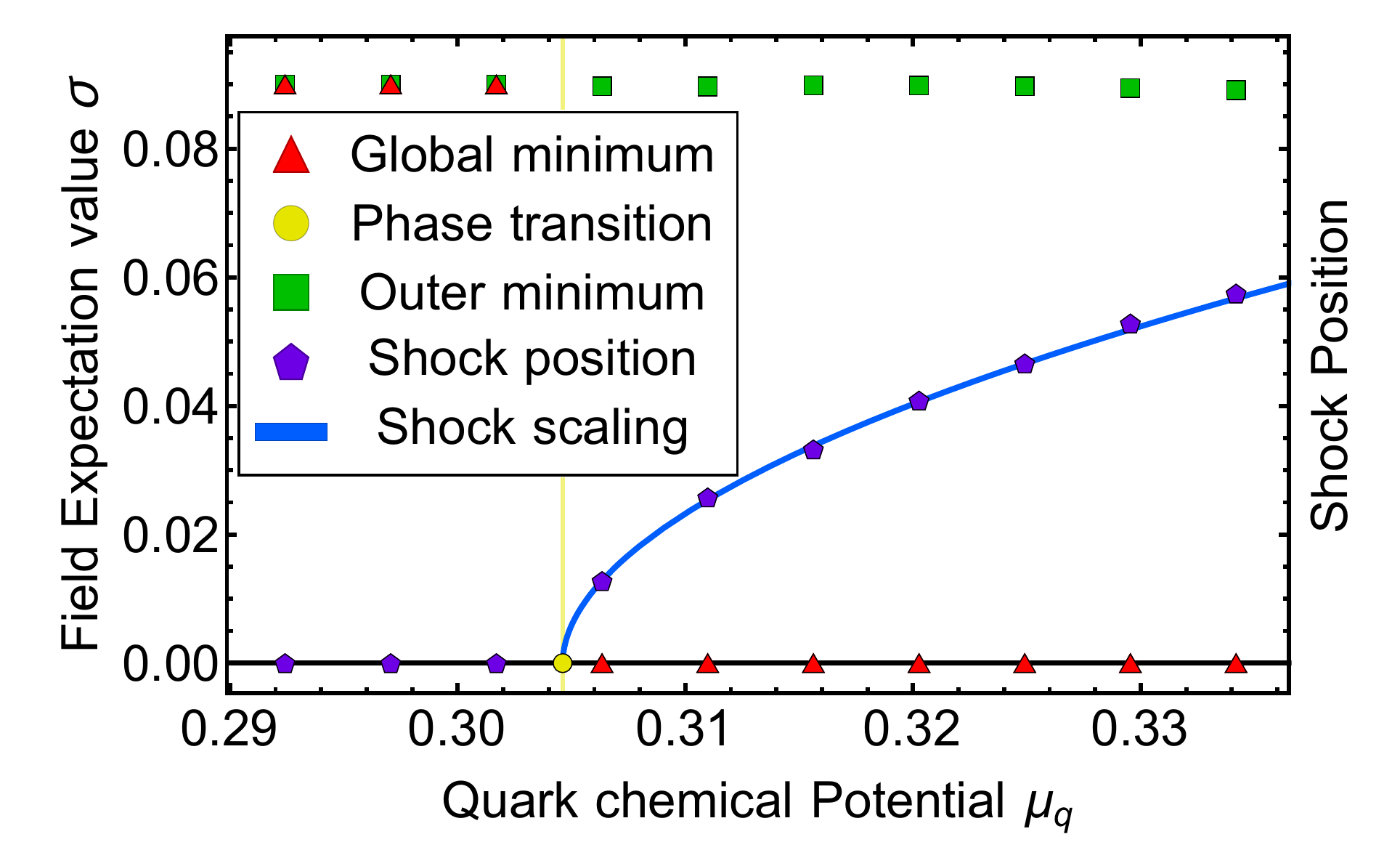} 
	\caption{Density dependence of the mesonic field expectation value $\sigma_{0}$ at $T\approx 0$ for the large-$N_f$ model  with 3 DoFs in LPA (constant Yukawa coupling). The figure shows a first order phase transition of the field expectation value at zero temperature. The solution now contains a local outer minimum and shock development between the outer local and inner global minimum. The extrapolated shock position at $k=0$\,GeV is plotted at different densities. We find a second order phase transition of the order parameter. The parameters of a scaling fit, see \labelcref{eq:fit}, are given in \Cref{tab:fit}.\hspace*{\fill} }
	\label{fig:shock}
\end{figure}

Naturally, the occurrence of shock development depends on the choice of initial conditions, specifically those that a trigger stronger dynamics of the system, for a respective discussion in the $O(1)$-model see \cite{Grossi:2019urj}. With physical initial conditions we find shock development in the large $N_f$ limit with 3 DoF, whereas the dynamics for 4 DoFs and finite $N_f$ are not strong enough to generate a shock at finite temperature $T>10$ MeV. This is an important observation: we have used the same initial conditions for all models, fixed within the two-flavour case. As discussed before, the two models in the large $N_f$-limit only differ by a rescaling of the fields and parameters. Accordingly, they can be interpreted as the same model with different initial conditions, as we do not apply any rescaling to the initial condition. However, these changes are marginal, as can be seen from the small variation of the pion decay constants and $\sigma$-masses in the vacuum, see \Cref{tab:rho0}. In conclusion the physical case is very close to the situation where shocks may form during the RG-time evolution. Whether or not this also occurs in QCD requires further investigation: 
\begin{itemize}
	\item [(i)] The embedding of the present model as part of the matter sector will lead to additional driving forces in the flow. This may be mimicked with a $T,\mu$-dependent change of the initial conditions here. naturally these changes can go either way, they may support the shock development or soften it.  
	\item [(ii)] The additional diffusion terms in the finite-$N_f$ case, see \Cref{sec:FiniteN}, may structurally soften the RG-time evolution and remove any shock development. It is also unclear whether shocks develop in the presence of  diffusion terms allow, as the diffusive flux counteracts the formation of a discontinuity.
\end{itemize}
The resolution of these aspects is crucial for an access to the QCD phase boundary at large chemical potential and low temperatures. This goes far beyond the scopes of the present work, and is subject of ongoing work. 

\begin{table}[t] 
	\vspace{.2cm}
	\centering
	\begin{tabular}{|p{0.18\linewidth}||p{0.18\linewidth}|p{0.25\linewidth}|p{0.25\linewidth}|}
		\hline 
		& & & \\[-1ex]
		Parameter& Prefactor & Crit. exponent & Crit. density \\[1ex]
		& $\beta$ & $\zeta$ & $\mu_{\mathrm{crit}}$ \\[1ex]
		\hline \hline
		& & & \\[-1ex]
		$\chi^2$-Fit & 0.391(45) & 0.524(31) & 0.30460(33) \\[1ex]
		\hline
	\end{tabular}
	\caption{Parameters obtained from a $\chi^2$-fit of \labelcref{eq:fit} to the shock positions in \Cref{fig:shock}.\hspace*{\fill} }
	\label{tab:fit}	
\end{table}
In the large $N_f$ limit with 3 DoFs we can use the shock development at low temperatures for an accurate determination of the phase transition line. The shock position $\xi_{\mathrm{final}}$ at $k=0$ is extrapolated by 
fit, utilizing the exponential decay of the flows, 
\begin{align*}
\xi(t)= \xi_{\mathrm{final}} + const \ e^{-t} \,.
\end{align*}
We use the shock position at 6 equally spaced time steps between RG-times $t=3$ and $t=3.5$. We expect the same power law behavior as in \cite{Grossi:2019urj} for the final shock position as a function of chemical potential, 
\begin{align}
\xi_{\mathrm{final}} = 
\begin{cases}
\beta |\mu - \mu_{\mathrm{crit}}|^\zeta, & \mu \geq \mu_{\mathrm{crit}} \\
0, & \mu \le \mu_{\mathrm{crit}}
\end{cases} \,.
\label{eq:fit}
\end{align}
As an explicit example we concentrate on $T = 10\,$MeV. 
The coefficients of the $\chi^2$-fit are provided in \Cref{tab:fit}. From this second order phase transition we obtain an accurate estimate for the critical chemical potential of $\mu_{\mathrm{crit}} = 0.30460 \pm 0.00033\,$GeV. 
\begin{figure*}[ht!]
	\begin{minipage}[b]{0.49\linewidth}
		\includegraphics[width=\linewidth]{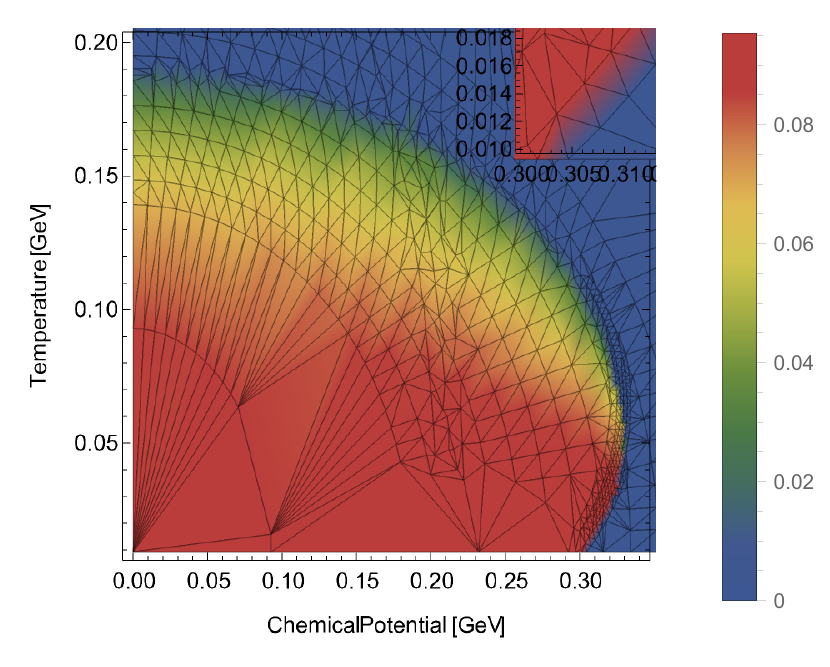}
		\subcaption{Large $N_f$ with 3 DoF, with slight shock development at high densities.\hspace*{\fill} }
		\label{fig:ln3}
	\end{minipage}\hspace{2mm}
	\begin{minipage}[b]{0.49\linewidth}
		\includegraphics[width=\linewidth]{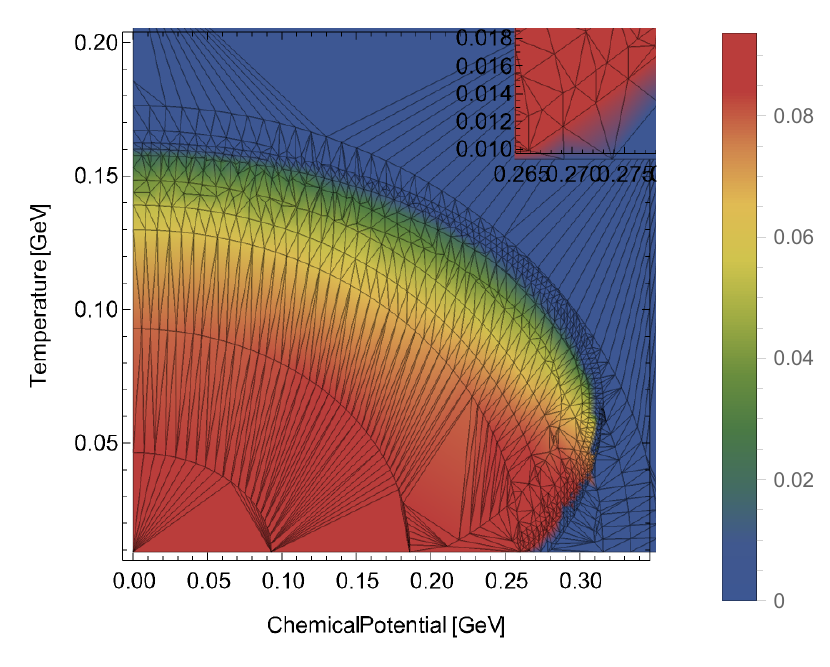}
		\subcaption{Large $N_f$ with 4 DoF and and no shock development at high densities.\hspace*{\fill} }
		\label{fig:dynamic}
	\end{minipage}	
	\caption{Phase diagram of the large-N model in the $(T,\mu)$ plane with constant Yukawa coupling, where the mean field expectation value $\sigma$ is chosen as order parameter. The mesh indicates the discrete data points. \hspace*{\fill} }
	\label{fig:pd}	
\end{figure*}
The phase transition and shock positions are depicted in \Cref{fig:shock}. Shock formation occurs only in a relatively small area of the $(T,\mu)$ plane, being confined to intermediate densities $290 \  \unit{MeV} < 360$\,MeV and small temperatures up to $T=20$\,MeV. 

In the absence of a shock a very fine grid has to be used to pin down the phase transition line for low temperatures. In the present work we have simply narrowed down the location of the phase transition line for small temperatures $T\lesssim 30\,$MeV to a small interval $\mu_{\mathrm{crit}} \in [270,290] $\,MeV. 

\begin{figure}[b]
	\includegraphics[width=\linewidth]{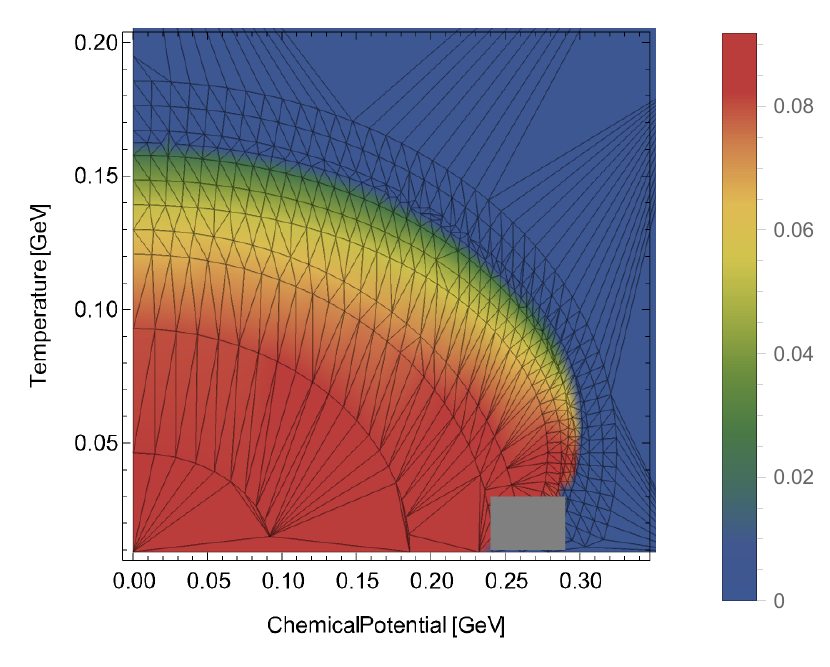}
	\caption{Phase diagram at $N_f=2$ with constant Yukawa coupling. The gray box indicates the points that did not converge. A thorough discussion of the convergence issues is provided in \Cref{app:conv-qm}. \hspace*{\fill} }
	\label{fig:pd-qm}
\end{figure}
\begin{figure*}[t]
	\centering
	\begin{minipage}[b]{0.49\linewidth}
		\centering
		\includegraphics[width=\linewidth]{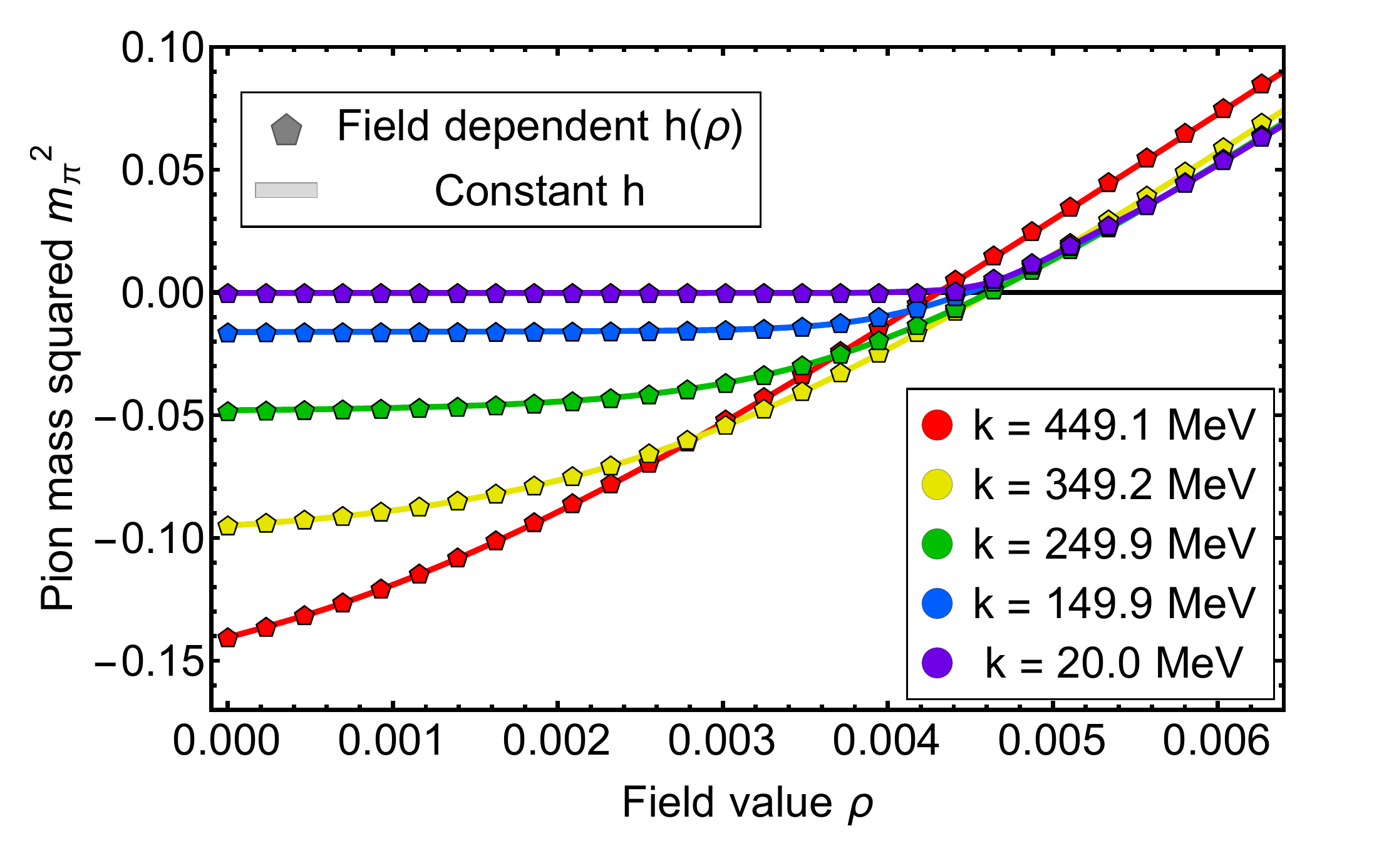}
		\subcaption{Comparision of the RG-running of the pion mass with constant and field dependent Yukawa coupling.\hspace*{\fill}}
		\label{fig:ypot}
	\end{minipage}%
	\hspace{0.01\linewidth}
	\begin{minipage}[b]{0.49\linewidth}
		\centering
		\includegraphics[width=\linewidth]{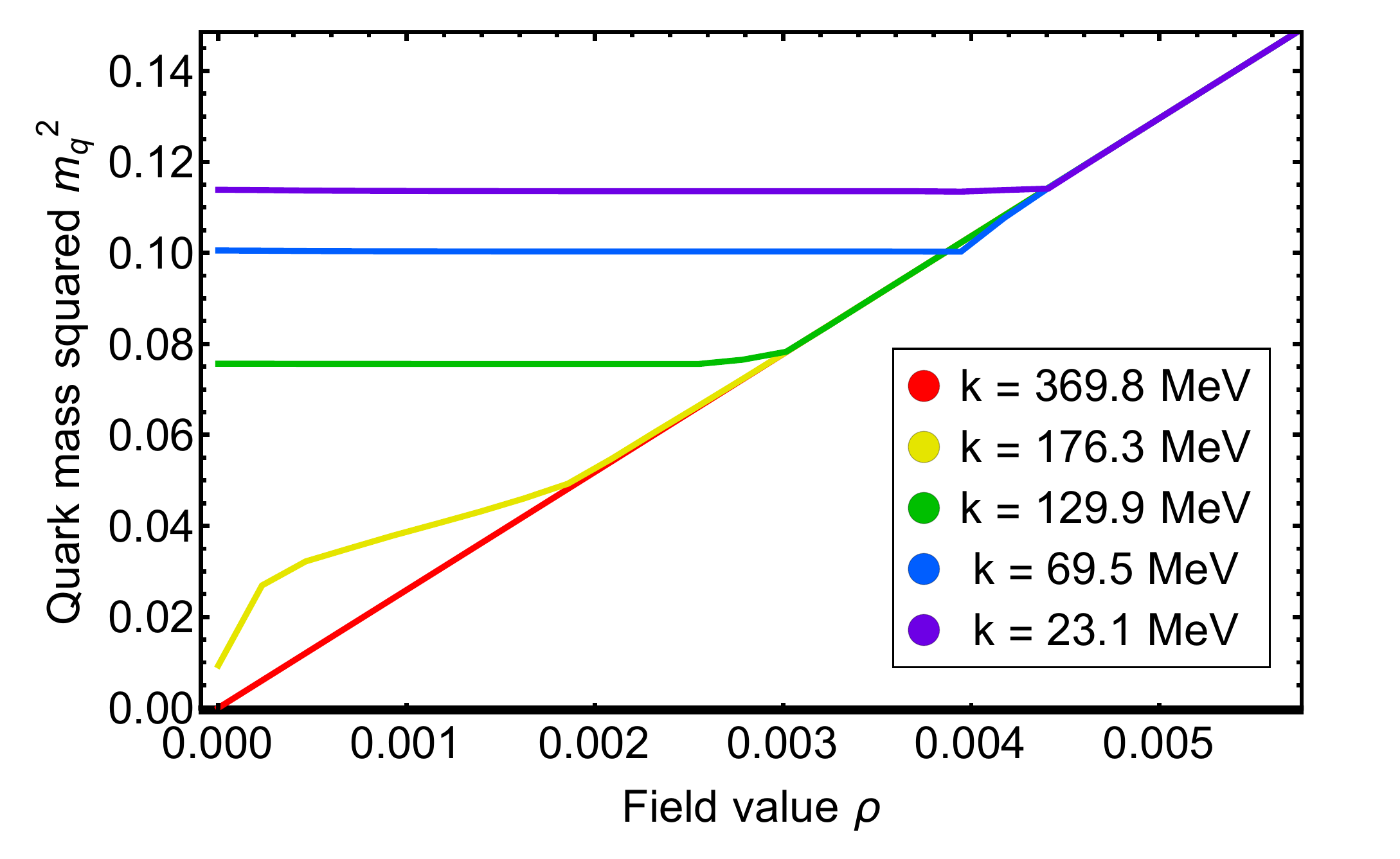}
		\subcaption{Field-dependence of the quark mass. A flat regime emerges for $k\to 0$, related to that in the effective potenial.\hspace*{\fill} }
		\label{fig:yqmass}
	\end{minipage}
	\caption{RG-scale evolution of the field-dependent quark and pion masses in approximate vacuum in the large-$N_f$ limit with 4 DoF. \hspace*{\fill}  }
	\label{fig:yuk-overview}
\end{figure*}

\subsubsection{Phase Structure in LPA}\label{sec:PhaseLPA}

The preparations and results discussed in the last sections allow us to compute the phase structure of the QM-model, both for $N_f=2$ and in the large-$N_f$ limit. 

In these computations we use a grid $\rho=[0, 0.03]$ GeV and expand in $K=70$ elements with a local approximation order of $N_p=2$. This setup ensure convergence of the numerics for all temperatures up to $T= 0.3$\,GeV and chemical potentials around $\mu= 0.35$\,GeV. 

For the resolution of shock formation a finer grid is required. We observe a formation of shocks around the first order phase transition in the large-N case at densities around $\mu = 0.3$\,GeV. In this area we expand in $K=250$ cells to reduce oscillations and ensure numerical convergence. The flow is evaluated up to $t=4$, which corresponds to a momentum scale of $k = 0.001$\,GeV. The field expectation value $\sigma$ is chosen as order parameter and evaluated as demonstrated previously for the approximate vacuum.

The result for the large-N limit is depicted in \Cref{fig:pd}. The crossover region is discernible by the color gradient that smoothly transitions between both phases, whereas in the first order regime a jump is clearly visible. \Cref{fig:dynamic} also illustrates how shock development shifts the critical chemical potential to higher values. 

The phase diagram for the  finite $N_f$ case is given in \Cref{fig:pd-qm}, the computation did not converge at high densities, this is further discussed in \Cref{app:conv-qm}. As discussed before, it will be interesting to see, how this regime changes, if the present model is embedded as part of the matter sector of QCD in full QCD-flows. This is subject of ongoing work. 
\subsection{Quark-meson scatterings in the Large-N limit}\label{sec:resy} 

The discussion of the results in LPA with a constant Yukawa coupling have revealed a very intricate structure at about onset chemical potentials and low temperatures. In particular the occurrence of shocks is very sensitive to the details of the dynamics. Moreover, we expect quark-meson scattering also being important in the vicinity of a potential critical end point in QCD.  

We now present our results for the QM-model with a field-dependent Yukawa coupling in the large-$N_f$ limit with four DoFs, based on the combined numerical solution of the flows \labelcref{eq:udrho} and \labelcref{eq:hdrho}, as formulated in \labelcref{eq:fin}.

\subsubsection{Dynamics in the vacuum}\label{sec:DynVac}

For the discussion of the dynamics in the vacuum, \labelcref{eq:fin} is solved on a grid with varying cell sizes. A local approximation order of $N_p = 3$ is used with $K=100$ cells in $\rho \in [0,0.02]$. \Cref{fig:yuk-overview} depicts the solutions of the pion and quark masses in approximate vacuum, the pion mass in comparison to the case with constant Yukawa coupling. We can see that in approximate vacuum the pion mass remains unchanged for both models. An exponential fit is performed on the position of the zero point of $\partial_\rho V_k(\rho)$ using 5 equidistant RG-scales from $k=65$ MeV to $k=25$ MeV and we obtain 
\begin{align}\label{eq:sigma0full}
\sigma_{0,\pi} = 87.4(17) \,\unit{MeV} \,. 
\end{align} 
This is consistent with the results in the previous sections with a constant Yukawa coupling (LPA). Consequently, this confirms previous findings in \cite{Pawlowski:2014zaa}, that LPA or rather higher orders in the derivative expansion are a good approximation for vacuum QCD. 

\Cref{fig:yqmass} depicts the field-dependent quark mass $m_{q,k}^2(\rho)$ for different RG-times. As argued in \cite{Pawlowski:2014zaa}, the quark mass flattens for meson fields $\rho\leq \rho_0$. The computation in the present work puts these conceptual and preliminary numerical findings on a sound numerical footing. In summary, at vanishing cutoff scale $k=0$, this leaves us with a field-dependent quark mass $m_q^2(\rho)$ with 
\begin{align}
	\label{eq:QuarkMassMin}
	m_q^2(\rho) \geq m_q^2(\rho_0)\,. 
\end{align}
Note that while conceptually the field value $\rho_{0,q}$, below which the mass function flattens, has to agree with $\rho_0$, the solution of the EoM, numerically this is not fully guaranteed. Hence, this provides a further consistency check of the present scheme. For performing the respective reliability check, we have determined the position of the kink by subtracting $-\rho$ from the solution and taking the minimum. An exponential fit gives,  
\begin{align}\label{eq:sigma0q}
\sigma_{0,q} = 86.0(17)\,\unit{MeV} \,,
\end{align}
which coincides within its error $\sigma_{0}$ in \labelcref{eq:sigma0full}. The error is computed from the grid resolution and the error to the fit parameters. 

The quark mass in the flattened area in \Cref{fig:yqmass} is computed as follows: one computes the average value of the quark mass up to the kink for the previously mentioned 5 RG-scales and performs an exponential fit. This leads to the physical quark mass, 
\begin{align}\label{eq:mqfull}
m_q(\rho_0) = 309.635(85)\ \unit{MeV} \,.
\end{align}
The linear $\rho$-dependence of $m_q^2(\rho)$ in \Cref{fig:yqmass} for $\rho> \rho_0$ entails, that the Yukawa coupling is constant with $\partial_\rho h(\rho) \approx 0$, already observed in \cite{Pawlowski:2014zaa} for finite $N_f$. The constant approximation for these field values works so well, that we can use the input Yukawa coupling $h_\Lambda$ to confirm $\sigma_0$ with the consistency  relation $\sigma_0 =m_q(\sigma_0)/h_\Lambda$. This leads us to $\sigma_{0,q'} = 86.01(24)\,\unit{MeV}$, where we take the error from the fit and the mean deviation in the flattened region. 

In summary the present numerical analysis confirms quantitatively the conceptual and preliminary numerical analysis in \cite{Pawlowski:2014zaa}: In the broken phase the flattening-out of the field-dependent quark-mass $m_q^2(\rho)$ is triggered by the flattening of the effective potential. In turn, in the symmetric phase the field dependent quark mass does not flatten-out and the quark mass vanishes on the solution of the equation of motion, $\rho_0=0$. Respective plots of the field-dependent quark mass $m_q^2(\rho)$ and the pion mass $m_\pi^2(\rho)$ for high temperature and density values are discussed in \Cref{app:resyuk}. Importantly, apart from the flattening of the quark mass, they do not deviate significantly from the results in LPA. In particular this applies to their values of the equations of motion. Note however, that this is bound to change for finite $N_f$. 

\subsubsection{Phase structure} \label{sec:PhaseFull}

As in LPA with a constant Yukawa coupling, we finally present our results on the phase diagram of the QM-model in the large-$N_f$ limit including quark-meson scatterings via a field-dependent quark mass or Yukawa coupling. 

\begin{figure}[t!]
	\includegraphics[width=\linewidth]{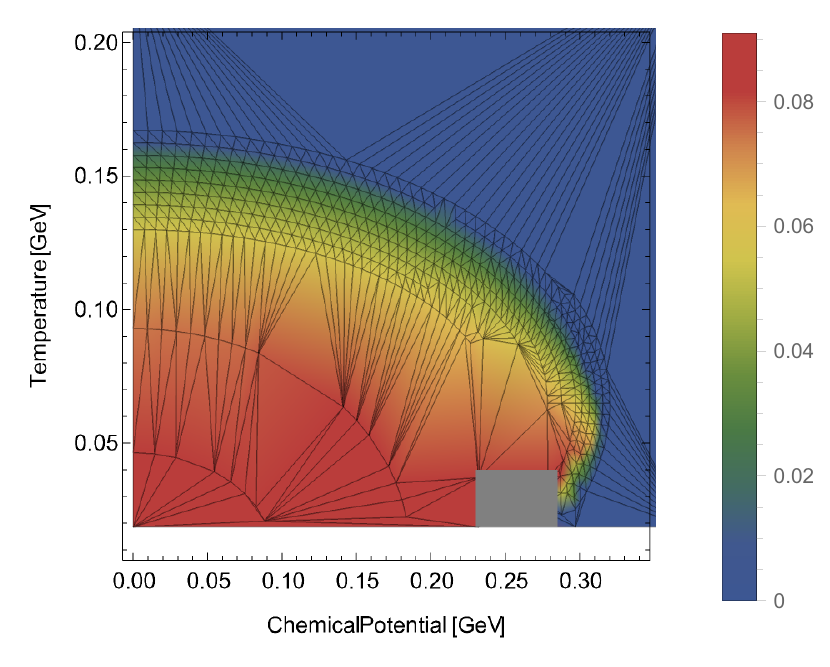}
	\caption{Phase diagram for the quark-meson model in the large-N limit with four DoFs with field dependent Yukawa coupling. The gray box indicates the points that did not converge. The mesh indicates the discrete data points.\hspace*{\fill} }
	\label{fig:yphase}
\end{figure}

The computation uses the same resolution as above: $N_p = 2$ and $K=80$ cells in $\rho \in [0,0.02]$ up to the RG-time $t=4$, that is $k=0.001$\,GeV. The result is shown in \Cref{fig:yphase}. The computations did not complete the time integration for  $\mu \ge 0.3$ GeV, respective upgrades are currently investigated. 

As already discussed in the last section, $\sec${sec:DynVac}, in the phase with chiral symmetry breaking the field-dependent quark-mass $m_q^2$ is necessarily flattened for $\rho\leq \rho_0$. In turn, The quark mass function does not flatten in the symmetric phase, and the quark mass is found to be zero in the symmetric phase. Plots of the field dependent quark mass $m_q^2$ ang pion mass $m_\pi^2$ for high values of temperature and density can be found in \Cref{app:resyuk} and do not significantly deviate from the results with a constant Yukawa coupling. 

\begin{figure}[t!]
	\includegraphics[width=\linewidth]{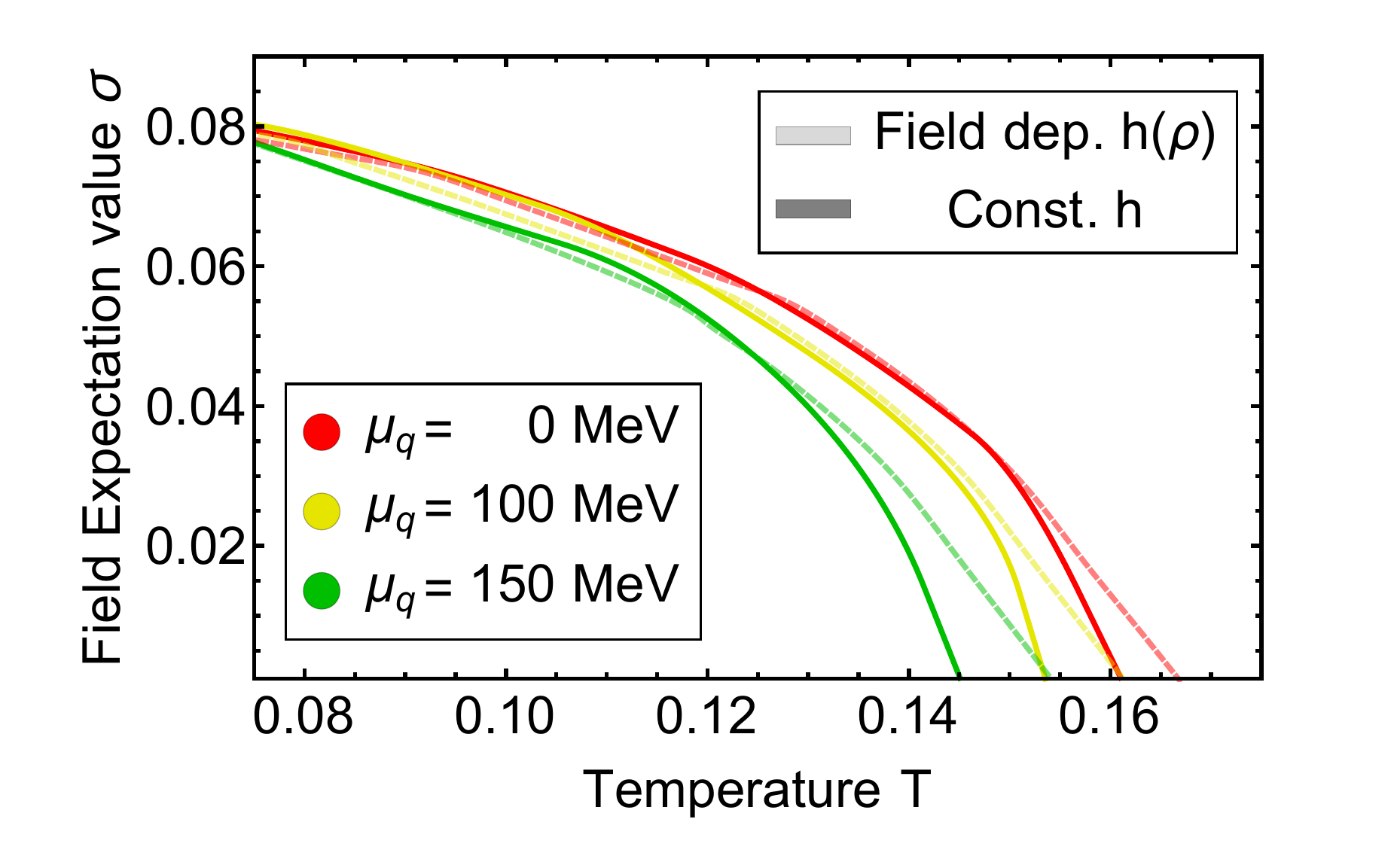}
	\caption{Chiral order parameter $\sigma_0(T)$ in the large-$N_f$ limit with four DoFs as a function of temperature and quark chemical potential. We compare the LPA results with quark-meson scatterings encoded in $m_q(\sigma)$ for different densities. The data is interpolated from the phase diagrams in \Cref{fig:pd} and \Cref{fig:yphase}.\hspace*{\fill} }
	\label{fig:ChiralO}
\end{figure}

We close this section with a comparison of the phase structure in \Cref{fig:yphase} with that in LPA, \Cref{fig:dynamic}, in the same setting: large-$N_f$ limit with four DoFs. While the phase boundaries do not change significantly, the crossover gets softened, if quark-meson scatterings are taken into account. This is clearly visible in \Cref{fig:ChiralO}, where we depict the chiral order parameter as a function of temperature for different densities with $\mu_q = 0, 100, 150$\,MeV. This entails, that the quark-meson scatterings considered here give sizable contributions to important observables measured in heavy ion collisions. First of all, fluctuation observables will be sensitive to such a widening of the crossover. These effects may be even more prominent at large chemical potential where the freeze-out line most probably deviates from the chiral crossover line. Moreover, it can also be deduced from \Cref{fig:ChiralO}, that the quark-meson scatterings may have a sizable delaying effect on a possible critical end point. We conclude, that these scatterings have to be taken into account for a quantitative prediction for the location of the CEP. 

%


\section{Conclusions}
\label{sec:sum}

We have presented a study of the QM-model with an emphasis on a quantitative access to order parameter potentials at finite chemical potentials. This allows us to discuss the eminently important question of the location of phase transition lines, that of the symmetry breaking pattern and the order of the phase transitions. 

The present study combines two systematic advances in the past years: The first one was the development of self-consistent approximations for the computation of order parameter potentials,  \cite{Pawlowski:2014zaa}. The second one was the development of a numerical approach for solving flow equations that also enables us to discuss discontinuities in the flows such as shocks that are potentially relevant for the correct description of first and second order phase transitions, \cite{Grossi:2019urj}. 

Within this approach we have computed the phase structure of the quark-meson model (QM-model) at finite temperature and density. An important benchmark is already provided in the large $N_f$-limit with an infinite number of flavours. We have argued that within an 't Hooft-type limit we can mimic the two-flavour QM-model well (or any other flavour), and in particular reproduce well its non-universal properties such as the location of the phase boundary. 

Moreover, in this limit the numerical approach within the discontinuous Galerkin set-up in \cite{Grossi:2019urj} is fully developed and we have a quantitative access to the shock-development and propagation, even in the presence of non-conservative forces. 

The present approach works very well except of a small regime at low temperature and onset densities. This is a merely technical problem and related upgrades of the present schemes are in development. Moreover, already for smaller ratios $\mu_q/T$ close to the crossover line we have to also improve the current approximation of the matter sector of QCD. This follows already from \cite{Fu:2019hdw, Braun:2019aow}. The results there indicate the potential relevance of non-trivial meson dispersions as well as the diquark channel at larger chemical potentials, $\mu_q/T \gtrsim 4/3$. Moreover, in the vicinity of a potential critical end point we also have to take into account the density channel, that mixed with the critical $\sigma$-mode. This is work in progress and we hope to report on it in the near future. 


\begin{acknowledgments}
We thank L.~Corell, L.~Kades, A~Koenigstein, F.~Rennecke, M.~Steil and J.~Urban for discussions. This work is done within the fQCD collaboration \cite{fQCD}, and is supported by EMMI, the BMBF grant 05P18VHFCA, and is part of and supported by the DFG Collaborative Research Centre SFB 1225 (ISOQUANT) as well as by the DFG under Germany's Excellence Strategy EXC - 2181/1 - 390900948 (the Heidelberg Excellence Cluster STRUCTURES).

\end{acknowledgments}


\appendix
\begingroup
\allowdisplaybreaks

		
\section{Implementation of DG-Methods in Dune} \label{app:numerics}
This section gives an introduction to the numerical and computational framework used to solve the equations \labelcref{eq:pdeform}. We made use of the DUNE (Distributed and Unified Numerics Environment) library \cite{dune:06,dum:11,dune:pdelab,dune-istl,dunepaperI:08,dunepaperII:08}, which is a modular toolbox for solving partial differential equations with grid-based methods. The library is an open source initiative to create a common interface for many different numerical methods and supports high performance computing . 

\subsection{Weak formulation and discrete problem}

The system of equations is solved on a computational domain $\Omega_h$, which is composed of $K$ disjoint elements, called cells, $D^k$ such that:
\begin{equation*}
\Omega \simeq \Omega_h = \bigcup\limits_{k=1}^{K} D^k \,.
\end{equation*}
For the purpose of the calculations in this work we used the Dune-grid \textit{YaspGrid}, which is contained in the module \textit{dune-grid} and allows for n-dimensional cubic grids and parallelised computation. In this paper a one dimensional grid is used. It represents the computational domain $\Omega_h$, with the grid-cells being disjoint intervals $D^k$ of possibly differing lengths, as discussed in \Cref{sec:galerkin}. In a more general formulation the domain $\Omega_h$ would be given as an n-dimensional rectangular grid and the elements $D^k$ would be implemented as cubic grid cells. 
	
The solution in each cell $D^k$ is approximated by 
\begin{equation*}
u(t,x) \simeq u_h(t,x) = \bigoplus\limits_{k=1}^{K} u_h^k(t,x) \,,
\end{equation*}
where $u_h^k(t,x) $ is the local solution in each cell and the index $h$ denotes the approximation. The local solution in turn is then approximated by a polynomial of degree $N = N_p -1$ such that in each element $ D^k$:
\begin{align}\label{eq:approx}
	u_h^k(t,x) = \sum\limits_{n=1}^{N_p} \hat{u}_n^k(t)q_n(x)\,. 	
\end{align}
The local approximation $u_h^k(t,x) $ is given by a modal expansion, where $\{q_n\}$ is a local polynomial basis with time dependent expansion coefficients $\hat{u}_n^k(t)$. Thus the global solution consists of $K$ local polynomial solutions of order $N$. The local approximaltion was implemented using the \textit{dune-pdelab} module, specifically using the class \textit{QkDGLocalFiniteElementMap}. In the one-dimensional case basis functions $q_n$ are given by the Legendre-Polynomials up to order $N_p$. For the purpose of higher dimensional computations the basis functions are taken from the polynomial space $Q_k$ of the Legendre-Polynomials. 
	
In our calculations we use a locally defined weak formulation of the convergence requirement:
\begin{align} \nonumber
	\int_{D^k} \Big( (\partial_t u_{i,h}+  a_{i,h}\partial_\rho u_{i,h} + s_{i,h})q_n  + f_{i,h} \partial_x q_n \Big)d x \\[1ex]
	= - \int_{ \partial D^k}  q_n  \Big(f_i^* \hat{\mathbf{n}} + \mathbf{D}(u_{i,h}^+, u_{i,h}^-, \hat{\mathbf{n}}) \Big) dx \,.
	\label{eq:weak}
\end{align}
The right hand side of the equation contains the standard numerical flux $f^*$ as well as an additional non-conservative flux term $\mathbf{D}$. $\hat{\mathbf{n}}$ is the outward pointing normal vector. We chose to use the local Lax-Friedrichs flux for $f^*$, which averages the flux on both sides of boundary and adds an additional diffusion term smoothing out jumps across the boundary:
\begin{align}
	f^* (u^+_h, u^-_h) = \frac{1}{2} \big(f_h(u_h^+)+f_h(u_h^-)\big) + \frac{C}{2} \mathbf{[[u_h]]}  \,,
	\label{eq:lax-fried}
\end{align}
where  and the indices + and - denote the outward (neighboring) and interior element at the boundary. The brackets denote a jump across the boundary, 
\begin{align*}
	\mathbf{[[u]]} = \big(\hat{\mathbf{n}}^-u^-  +\hat{\mathbf{n}}^+u^+ \big) \,.
\end{align*} 
$f_{max}$ is the local maximal wave speed, which corresponds to the speed of the fastest propagating mode across the boundary. In one dimensions this corresponds to
\begin{equation}
	f_{max} \geq \max\limits_{D^{\{ i, i+1 \}}} |\partial_u f(u)| \,.
	\label{eq:wavespeed}
\end{equation}
The local Lax-Friedrichs flux is the most natural extension from the analytic solution of linear conservation laws to the non-linear case. It relies on the so called Roe condition, which reflects the assumption that the system is dominated by one strong wave. 
%

\subsection{Non-conservative product}
The additional non-conservative flux across a boundary is given by $\mathbf{D}$. The theory of non-conservative fluxes was developed in \cite{pares2006numerical,castro2006high} and is applied in the context of Finite Volume and Discontinuous Galerkin schemes \cite{hou1994nonconservative,CASTRO2016347, dumbser2010force,dumbser2009ader, castro2010some, castro2008many, Dumbser2010, cf7315f2fac84dc9beed7e7b65f54ec3, PhysRevD.97.084053}.
To compute this quantity we need to consider the general form of a flux across an interface. For this purpose we consider a path $\phi_i(s)$ along the solution $u_i$, with start and endpoint $u_i^L$ and $u_i^R$ respectively and the parameter $s \in [0,1]$. The formal definition of the flux along this path for a non-conservative flux contribution $a_i \partial_\rho u_j$ (see \labelcref{eq:pdeform}) is then given by:
\begin{align}
	f_{i,\mathrm{nc}} = \int_{0}^{1} a_i (\phi_i(s), \phi_j(s),s)  \partial_s \phi_j(s) ds \,.
\end{align}
We remark that in the non-conservative case the flux is dependent on the chosen path. 
	
By choosing the right and left sides of a boundary $u^R=u^+$ and $u^L=u^-$ we are able to compute the flux from one cell to another. Similar to the numerical flux, $\mathbf{D}$ has to satisfy the jump property for consistency, 
\begin{align*}
	\mathbf{D}(u^+, u^-, \hat{\mathbf{n}}) +\mathbf{D}(u^+, u^-, -\hat{\mathbf{n}}) = \int_{0}^{1} a(\phi(s)) \hat{\mathbf{n}} \frac{\partial  \phi}{\partial s} ds \,,
\end{align*}
\begin{figure*}[t]
		\centering
		\begin{minipage}[b]{0.49\linewidth}
			\centering
			\includegraphics[width=8cm]{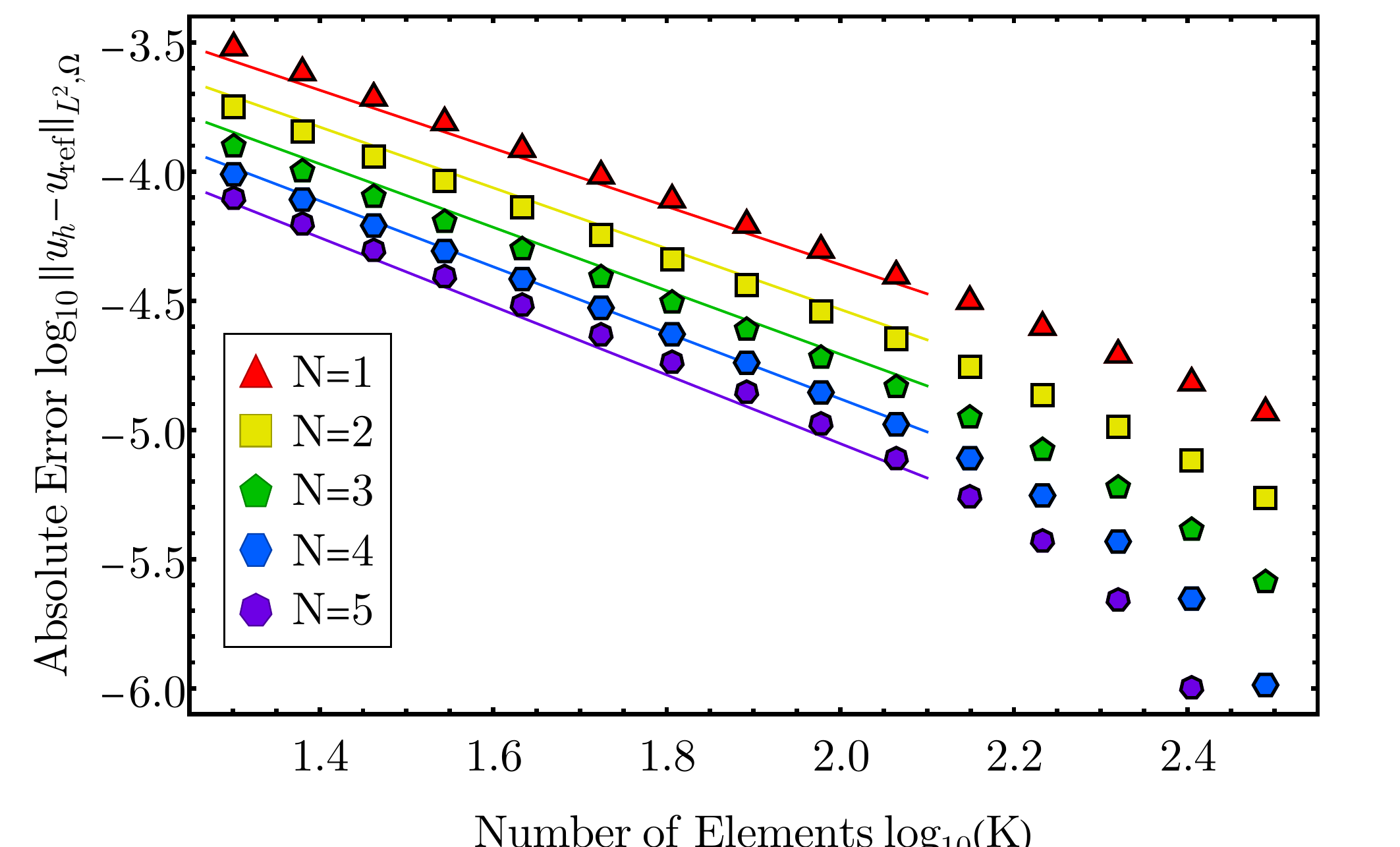}
			\subcaption{Error of the solution at $\mu = 0$ MeV. The lines show the $\chi^2$-fit of \Cref{app:fit} to all but the last 5 datapoints, with paramters given in \Cref{app:tab}.\hspace*{\fill} }
			\label{app:nochem}
		\end{minipage}%
	\hspace{1mm}
		\begin{minipage}[b]{0.49\linewidth}
			\centering
			\includegraphics[width=8cm]{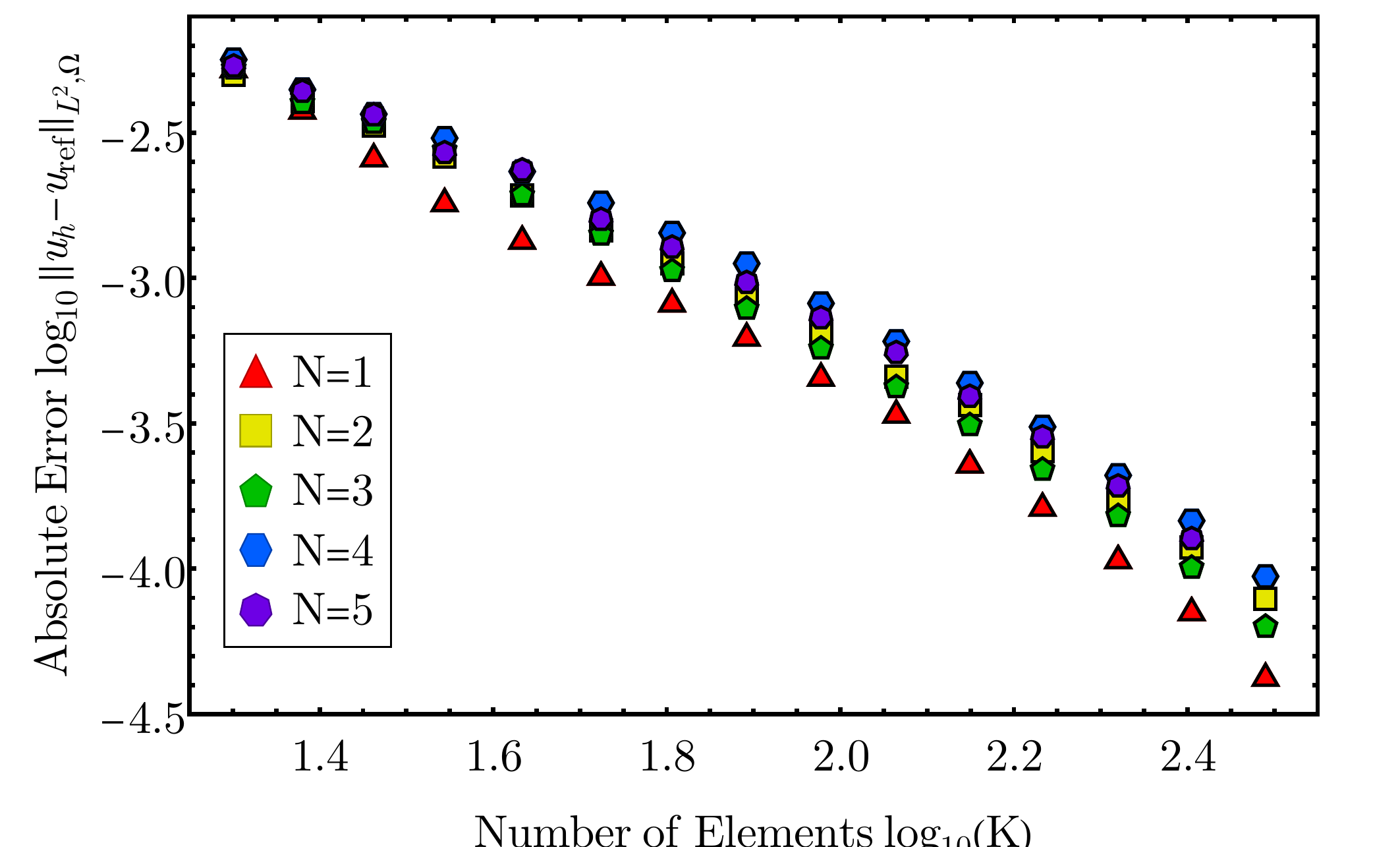}
			\subcaption{Error of the solution at $\mu = 400$ MeV.\hspace*{\fill} }
			\vspace{0.8cm}
			\label{app:chem}
		\end{minipage}
		\caption{Error of the solution with respect to the reference computation $u_\mathrm{ref}$ with $K=700$ and $N_p=5$ at $T=10$\,MeV and different values of the chemical potential. The solution was computed up to $k=150$\,MeV in an interval $0 \leq \rho \leq 0.02$ for different numbers of equally sized cells $K$ and local approximation order $N_p$. The symbols show the result of the numerical simulation.\hspace*{\fill} }
		\label{app:convergence}
\end{figure*}
which implies $D(u,u,\hat{\mathbf{n}})=0$ when there is no jump. 

This condition can be obtained by integrating the equation around a discontinuity. It generalizes the so called Ranking Huginot Condition for  non-conservative systems of equations. The numerical fluxes are, 
\begin{align}\label{eq:fullnoncons}
		\mathbf{D}(u^+, u^-, \hat{\mathbf{n}}) = &\frac{1}{2} \int_{0}^{1} a(\phi(s)) \hat{\mathbf{n}} \frac{\partial  \phi}{\partial s} ds \, \\
		&+\frac12  \int_{0}^{1} |a(\phi(s)) \hat{\mathbf{n}}| \frac{\partial  \phi}{\partial s} ds \ .
\end{align}
$|a(\phi(s)) \hat{\mathbf{n}}| $ is intended as the absolute value of the matrix namely $|a|= U^{-1 } \text{diag} (|\lambda_1|, \cdots, |\lambda_N|) U$, with $\lambda_i$ the eigenvalue of the matrix. It is possible to prove that this choice of flux reduces to the Lax-Friedrichs flux in the conservative case. If the dominant convection part of the equation is given by the conservative flux this extra term can be neglected. The non-conservative flux under this assumption can therefore be computed from 
\begin{align}\label{eq:noncons}
	\mathbf{D}(u^+, u^-, \hat{\mathbf{n}}) = \frac{1}{2} \int_{0}^{1} a(\phi(s)) \hat{\mathbf{n}} \frac{\partial  \phi}{\partial s} ds \,.
\end{align}
The last remaining degree of freedom are the boundary conditions for the outer boundary of $\Omega_h$. In our case they are given by the in- /out-flowing flux, which is implemented by setting $u_{i,h}^+ = u_{i,h}^-$ at the outer boundaries, effectively adding an imaginary additional cell. It follows that the non-conservative flux is not fit for flux-boundary conditions, since the non-conservative flux vanishes at the outer boundaries due to the jump property. Therefore the equations need to be reformulated such that the boundary-conditions can be met using the conservative flux. This is done in \Cref{app:non_cons_flux}.

\subsection{Time stepping}\label{sec:cfl}

The solution is computed by an explicit third-order time-stepping scheme from the \textit{dune-pdelab} module, where we additionally implemented Courant-Friedrichs-Lewy (CFL-)conditions. The time step $\Delta t$ is thus limited by the propagation speed of the flow:
	\begin{align}
	\Delta t \le \frac{\Delta x}{(2N +1)} \frac{1}{f_{max}} \,,
	\label{eq:cfl}
	\end{align}
where $\Delta x$ is the size of the grid cell, $N$ the polynomial degree used in the computation, such that the denominator indicates the total amount of grid points within the respective cell. $f_{max}$ is the maximal propagation speed of the information and was defined in \labelcref{eq:wavespeed}. 
	
Additionally we use a minmod slope limiter before each computation step to suppress oscillations around kinks and jumps in the solution. 

\section{Convergence Properties} \label{app:conv}

Results from the large-N computations with constant Yukawa coupling ( \Cref{sec:reslN}) of different approximation orders $N_p$  are compared to benchmark the accuracy of the computations. Since there is no analytic solution available to compute the numerical error generated by the DG-scheme, we use the result of a computation with $N_p=5$ and $K=700$ elements $u_{\mathrm{ref}}$ as a reference. The results are compared at an energy scale of $k= 140$\,MeV. The discrete time-stepping was adjusted such that the time-step computed by the CFL-condition is lowered to ensure $k=140$\,MeV is reached exactly. The discrete solution is used to generate an interpolating function $u$, from which the $L^2$ norm  
$
	|| u -u_{\mathrm{ref}}||_{L^2, \Omega_h} 
$,
is computed. This is done at temperature T=$10$\,MeV for $\mu = 0$\,MeV and $\mu = 400$\,MeV. The results are depicted in \Cref{app:nochem} and \Cref{app:chem} respectively. 
\begin{table}[b!] 
		\centering
		\begin{tabular}{c |c c c c}
			& & & &\\[-1ex]
			Parameter& $a_1$& $a_2$ & $b_1$ & $b_2$ \\[1ex] 
			\hline 
			& & & &\\[-1ex]
			$\chi^2$-Fit&2.016(80) & -0.067(25) & 1.088(45) & 0.054(14) \\[1ex]
			
		\end{tabular}
		\caption{Parameters obtained from a $\chi^2$-fit to the convergence behavior in \Cref{app:chem} using \Cref{app:fit}. }
		\label{app:tab}
\end{table} 
In vacuum we recover  good convergence properties observed in \cite{Grossi:2019urj} and we perform a fit of the first 10 data-points to the parametrization, 
\begin{align}
	\mathrm{log}_{10} & ||u-u_{\mathrm{ref}}||_{L^2, \Omega_h}\\[1ex]
	&= (a_1+a_2N_p)-(b_1+b_2N_p) \mathrm{log}_{10}(K) \,.
	\label{app:fit}
\end{align}
The convergence behaves like a power law when increasing the number of elements K and shows exponential convergence when increasing the local approximation order $N_p$ up to about $K=10^{2.1} \approx 125$. The fit parameters are given in \Cref{app:tab} and the fit to the data points is included in \Cref{app:nochem}. For higher $K$ we the error decreases even faster which is due to the numerical error of $u_{\mathrm{ref}}$ which e.g. also contains some of the resolution issues around the kink in the potential. When compared to \cite{Grossi:2019urj} the relative error between $u$ and the reference solution $u_\mathrm{ref}$ significantly bigger. This is in part caused by the fact that the functions were interpolated and not reconstructed using the original basis polynomials, since the Dune output only contains the values at cell boundaries with a precision of eight digits.
	
Adding a finite chemical potential increases the effect of the source term in \Cref{eq:udrho2}. At low temperatures the silver-blaze property introduces a sharp onset of chemical potential that shows better convergence for lower and odd $N_p$. However, since the exact solution is not known it is difficult to judge which local approximation is the best choice as a reference. \Cref{app:chem} illustrates the convergence properties in the area in which the source term dominates the equation over the flow. 
	
\subsection{Convergence in Systems with Diffusion}\label{app:conv-qm}
	
In this section we will comment on the Convergence of the equations in the Finite $N_f$ case.
\begin{itemize}
		\item We retain convergence properties similar to the previous section in regions of the phase diagram where the sigma mode is not critical. This is supported by the observation that the Finite-$N_f$ and the Large $N_f$ simulations behave very similarly (see eg. \Cref{fig:00}, \Cref{fig:30}).
		The Courant number is a factor $C$ chosen to ensure the inequality in \labelcref{eq:cfl}. In hydrodynamics $C=0.01$ is a common choice for diffusive systems, which is appropriate if the flow is not diffusion-dominated.
		\item In the diffusion dominated scenario, the solutions are not convergent. Additionally the time stepping becomes effectively 0. This is due to shock development in the solution, which creates a steep negative slope, i.e. a very high (divergent) diffusion flux contribution.	
\end{itemize} 
The lack of convergence is explained by the fact that approximate Riemann solvers are only applicable in convection dominated systems. The convergence of diffusion dominated flows can only be ensured using a new formulation of the fluxes, such as the local-DG methods \cite{doi:10.1137/S0036142997316712}.

\subsection{Convergence with a non Conservative Flux}

In this section we will comment on the convergence of the system with non conservative flux, i.e. the computations for field dependent Yukawa coupling. 
	
We can again distinguish two cases:
\begin{itemize}
		\item The area of the phase diagram where the pion mass $m_\pi^2$ has a single minimum for the entirety of the flow: In this case we retain similar convergence properties as in section \Cref{app:conv}. 
		\item At high chemical potential $\partial_\rho m_\pi^2$ begins to take on high negative values. This directly feeds back to the non-conservative flux. In this area the non-conservative flux dominates the flow, which is not contained in the CFL conditions.
	\end{itemize}
%

\section{Convexity restoration and time-stepping}\label{app:convex}

\begin{figure}[t!]
	\includegraphics[width=0.9\linewidth]{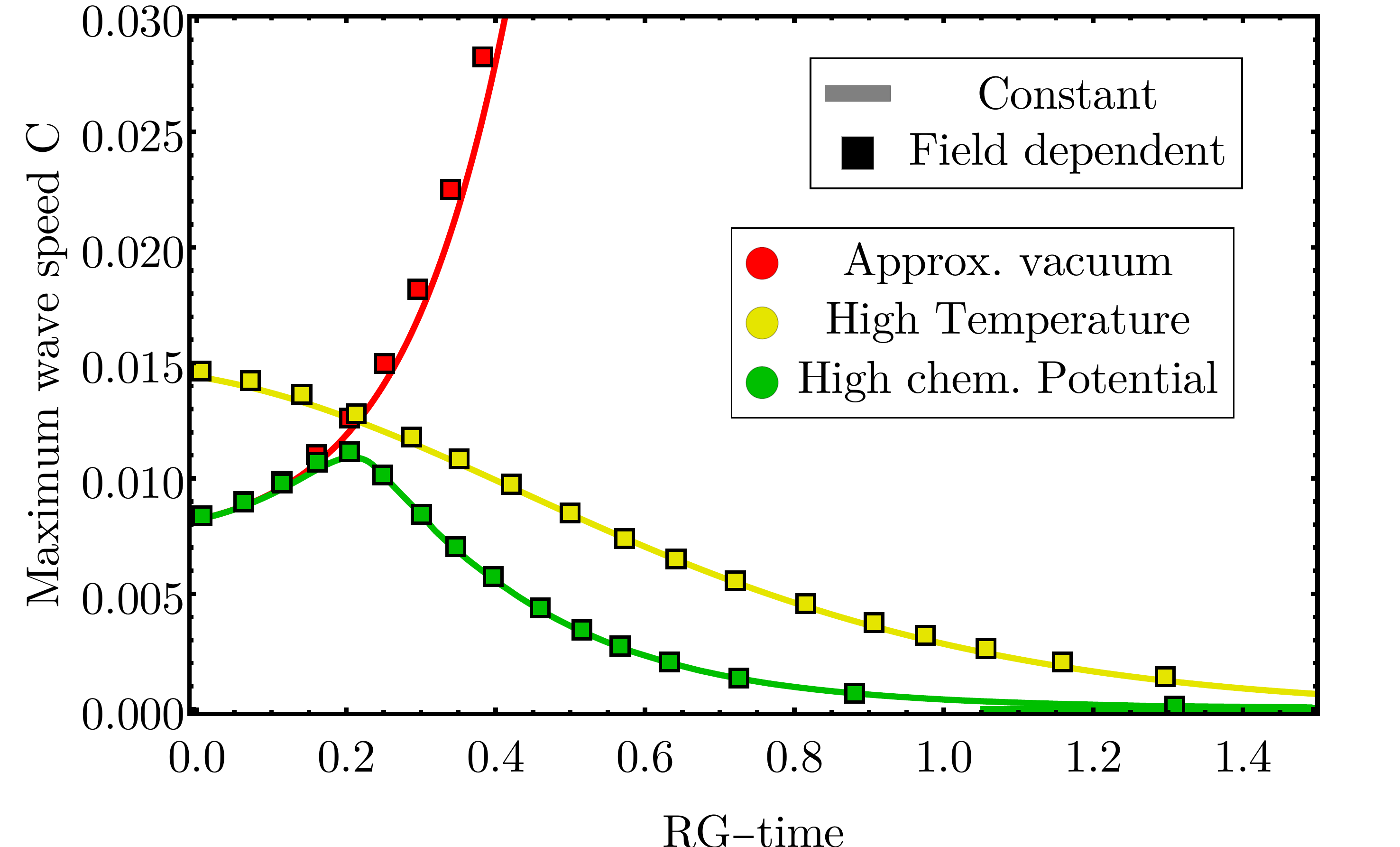}
	\caption{Comparison of the maximal information propagation speed $f_{max}$ for constant and field dependent Yukawa couplings at different places in the phase diagram. Both computations were performed with $K=100$ cells and a local approximation order $N_p=2$.\hspace*{\fill}  }
	\label{fig:wavespeed}
\end{figure}
In this section we are going to inspect the time-stepping and its related problems more closely. The equations are solved by a numerical step-wise integration of the RG-time using the CFL-conditions introduced in \Cref{sec:cfl}. The size of the integration step is dependent on the information flux between cells, the local wave speed $f_{max}$, which is defined in \labelcref{eq:wavespeed}. The local wave speed is plotted in \Cref{fig:wavespeed} for approximate vacuum in the broken symmetry phase and for high temperatures and chemical potential in the symmetric phase.

\subsection{Approaching Convexity in the Broken Symmetry Phase}

It can be observed from \Cref{fig:wavespeed} that the broken symmetry phase has a steadily increasing maximum wave speed. This corresponds to steadily decreasing time steps and leads to long computation times. This behavior is caused by the time-steps inverse proportionality to the flux:
\begin{align*}
\Delta t \propto (k^2 + u)^{n/2} \,,
\end{align*} 
for some positive integer value of $n$. The two-point function $\Gamma^{(2)}$ can have negative eigenvalues during the RG-flow, which is what happens to the computed function $u=m_\pi^2=\Gamma^{(2)}_{\pi \pi} $, the pion mass, in approximate vacuum. The flow is self-regulating, ensuring that the expression in the square root in the proportionality remains positive: The closer the root gets to becoming negative, the stronger the flow increases $u$, causing the modulo $|u|$ to always be slightly smaller than $k^2$. This must hold for $k \to 0$ from which it follows that $u \to 0$, such that $\Gamma^{(2)}=0$ at infinite RG-time and convexity is restored. $|u|$ teeters on the edge of becoming bigger than $k$ during the entire integration which results in a big flux between grid cells and very small time steps.

\subsection{Convexity in the Symmetric Phase}
The pion mass becomes positive at some point during the interpolation in the symmetric phase and convexity is restored before $k=0$. Positive values of $u$ also significantly decrease the flux between grid cells as can be seen from \Cref{fig:wavespeed} and increase the size of time steps.

At high temperatures the positiveness of $u$ is caused by the fact that the quark contribution to the flow that initially decreased $u$ is much smaller due to the dampening by the Fermi-Dirac Distribution. This translates into $k^2-u$ never being remotely close to $|0|$ and therefore no increased convexity ensuring flux. Initially the maximum wave speed at high chemical potentials is similar to the approximate vacuum. However, the sudden onset of density at $\ln{\big(\frac{\Gamma}{\mu}\big)}$ drives $u$ to positive values and the information flux decreases. 

It can be seen that the introduction of the field dependent Yukawa coupling only slightly increases the wave speed in the broken symmetry phase and has no effect in the symmetric phase, which is to be expected from the observation that $m_q^2$ barely changes during the RG-time evolution at high temperatures or densities made in \Cref{sec:resy}.

\subsection{Problems and Challenges with Time-Stepping}\label{app:restime}

A recurrent struggle while solving the non-linear RG-equations using CFL-conditions is time stepping. The RG-flow continuously works to restore convexity, relying heavily on the fact that $k^2 >|u|$ for negative $u$. For some cases where $k^2 >|u|$ is very small, a numerical error, for example an oscillation around a shock, can cause the radicant to become negative, and therefore no longer well defined. Limiting the time step by using \labelcref{eq:wavespeed} is spoiled further at high densities, where the sigma-mode becomes critical. The introduction of a steep slope in the potential introduces contributions to the flow that are not accounted for by time stepping.
\begin{figure*}[ht!]
	\centering
	\begin{minipage}[b]{0.49\linewidth}
		\centering
		\includegraphics[width=\linewidth]{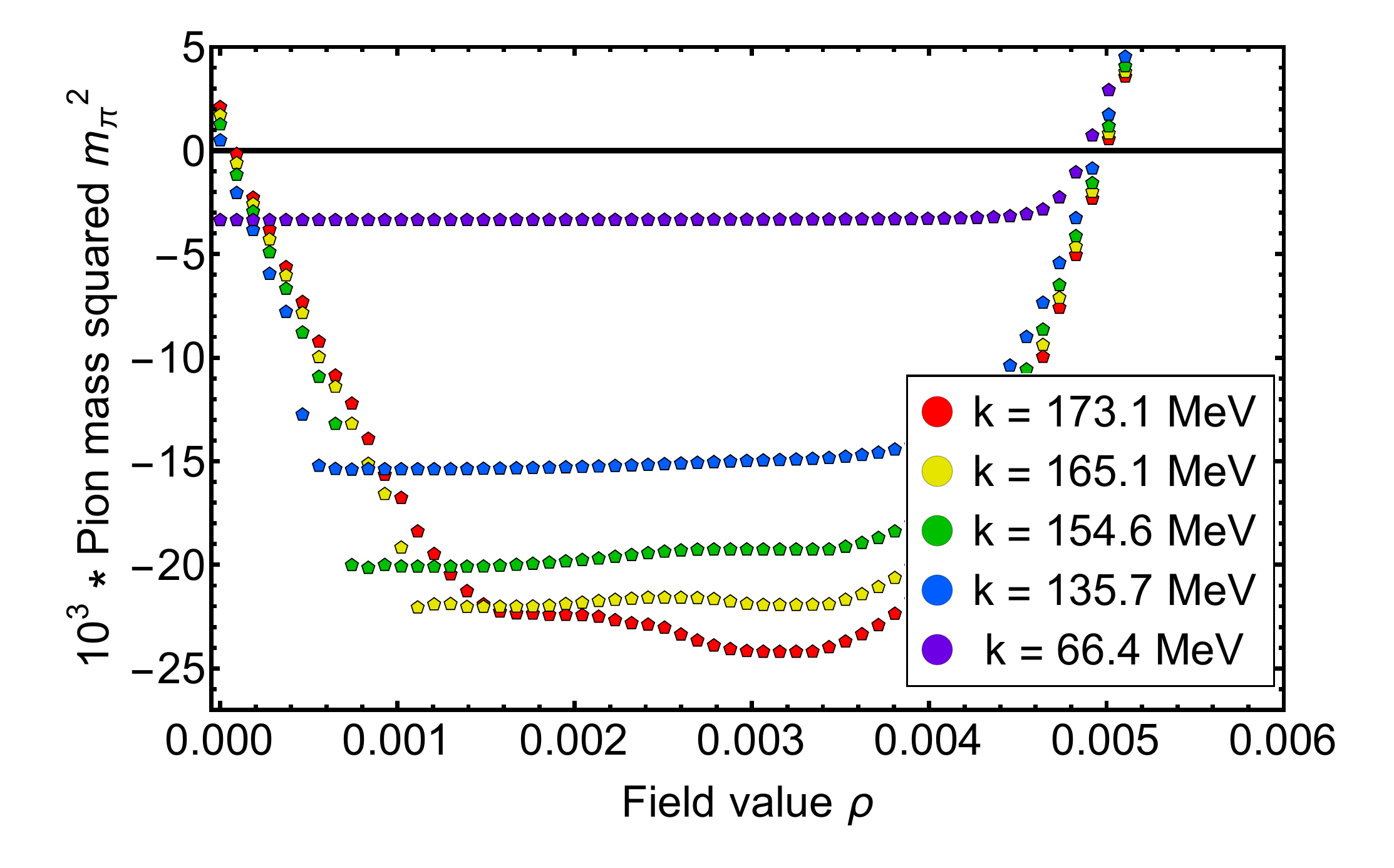}
		\subcaption{Derivative of the potential in the broken phase. The developing shock travels to unphysical values of the field value $\rho$ and the potential is flattened out.\hspace*{\fill} }
		\label{fig:s35}	
	\end{minipage}%
\hspace{1mm}
	\begin{minipage}[b]{0.49\linewidth}
		\centering
		\includegraphics[width=\linewidth]{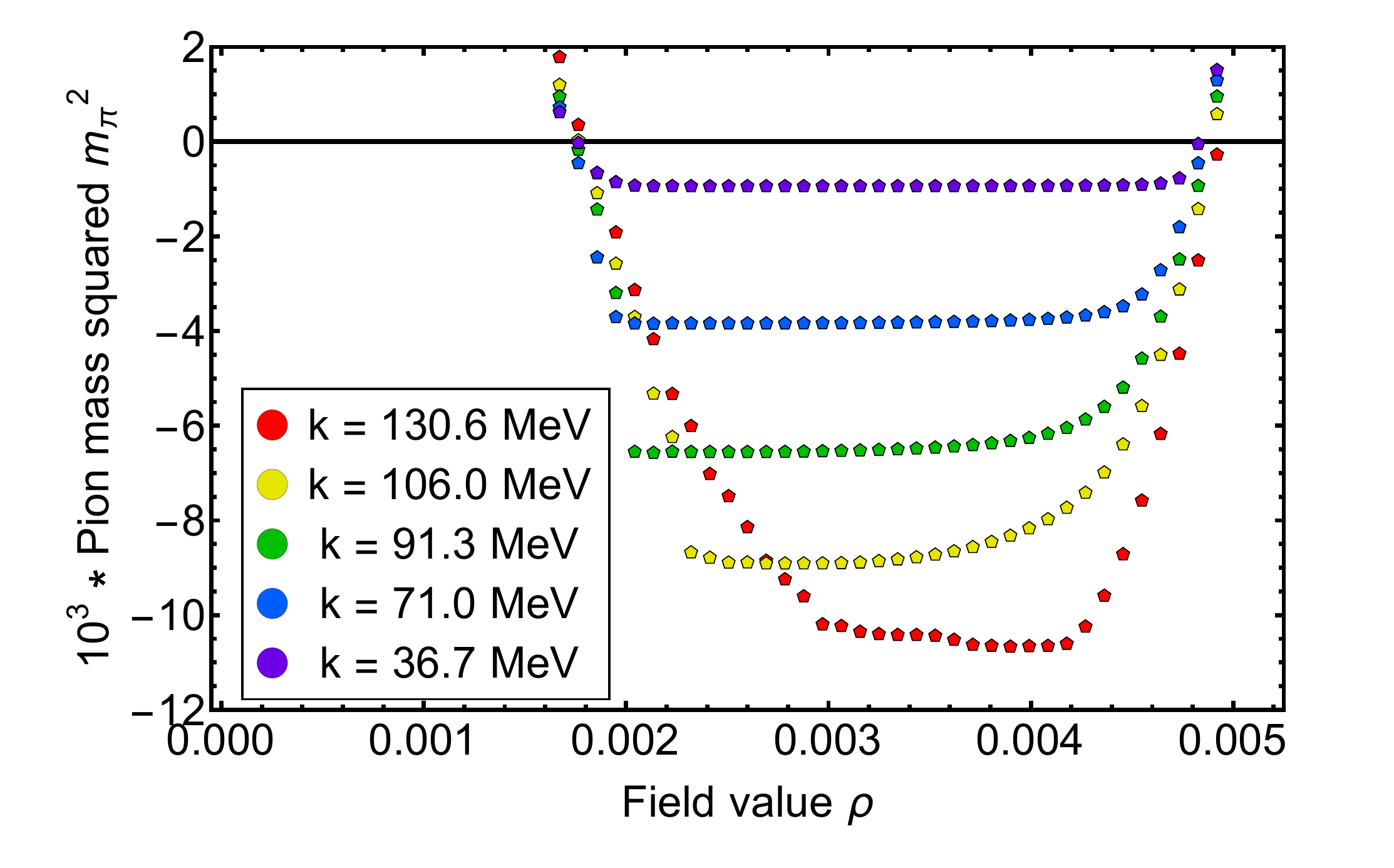}
		\subcaption{Derivative of the potential in the symmetric phase. The developing shock freezes in at finite $\rho$, creating a global mimumum at $\rho =0$.\hspace*{\fill} } \label{fig:s38}
	\end{minipage}	
	\caption{Shock development at high densities. The potential derivative $\partial_\rho V_k(\rho)$ is plotted in the vicinity of the phase transition at $\mu = 0.30$ GeV (\Cref{fig:s35}) and $\mu = 0.32$ GeV (\Cref{fig:s38}) in the large $N_f$ limit with $N_\pi=3$. The numerical oscillations around the shock were flattened out by a minmod slope limiter. The figure depicts the grid points on which the computation was carried out.\hspace*{\fill} }	
	\label{fig:shock2}
\end{figure*}
%
\section{Flow Equations of Pion and Quark Masses} \label{app:flow}

In this section the equations are reformulated to simplify their numerical treatment.
The flow equation of the pion mass is obtained by taking a $\rho$ derivative of the effective potential. In case of the large-N model this is given by, 
\begin{align} \label{eq:udrho2} \nonumber
\partial_t u_k^{\mathrm{lN}}(\rho)=&\, \partial_\rho \left[
\frac{k^5}{12 \pi^2} \left\{ \frac{N_\pi}{\epsilon_k^\pi} \left[1 + 2n_B(\epsilon_k^\pi)\right]\right.\right.\\[2ex]
& \hspace{-1.3cm}-\left.\left. \frac{ 4 \times 2 \times 3}{\epsilon_k^q}\left[1  -n_f(\epsilon_k^q + \mu)-n_f(\epsilon_k^q - \mu)\right] \right\}\right]
\,.	
\end{align}	

The flow equation of the Yukawa coupling in \labelcref{eq:hdrho} is rewritten in terms of the quark mass squared $m_q^2(\rho)$. To this aim, we multiply the original flow equation by $4 h(\rho) \rho$, which gives
\begin{align} \label{eq:mq2cons}
\begin{split}
\partial_tw_k =&\, 4 h_k \rho  A(u_k) \partial_\rho h_k  + 4 \rho h_k^4 B(w_k,u_k) \\[1ex]
=& 2 \rho A(u_k)\partial_\rho  h_k^2 + 4 \rho h_k^4 B(w_k,u_k) \\[1ex]
=&  A(u_k) \partial_\rho  w_k -2 h_k^2 A(u_k) + 4 \rho h_k^4 B(w_k,u_k)
\\[1ex]
=& A(u_k) \partial_\rho w_k+ \frac{w_k}{\rho} \Big[w_k B(w_k,u_k)-A(u_k)\Big]
\,,
\end{split}
\end{align}
where 
\begin{align} \label{eq:Ah}
A(m^2_{\pi,k};T,\mu) = -2 N_\pi v_3 k^2 l_1^{(B,4)}(m^2_{\pi,k};T) \,,
\end{align} 
corresponds to the contribution of the pion tadpole diagram and 
\begin{align} \label{eq:Bh}
B(m^2_{q,k},m^2_{\pi,k};T,\mu) = -4 N_\pi v_3 L^{(4)}_{(1,1)}(m^2_{q,k},m^2_{\pi,k};T,\mu) \,,
\end{align}
to the mixed contribution in \Cref{fig:h}. The explicit form of the threshold functions is given in \Cref{app:thfkt}. 
\begin{figure*}[t]
	\centering
	\begin{minipage}[b]{.49\linewidth}
		\includegraphics[width=\linewidth]{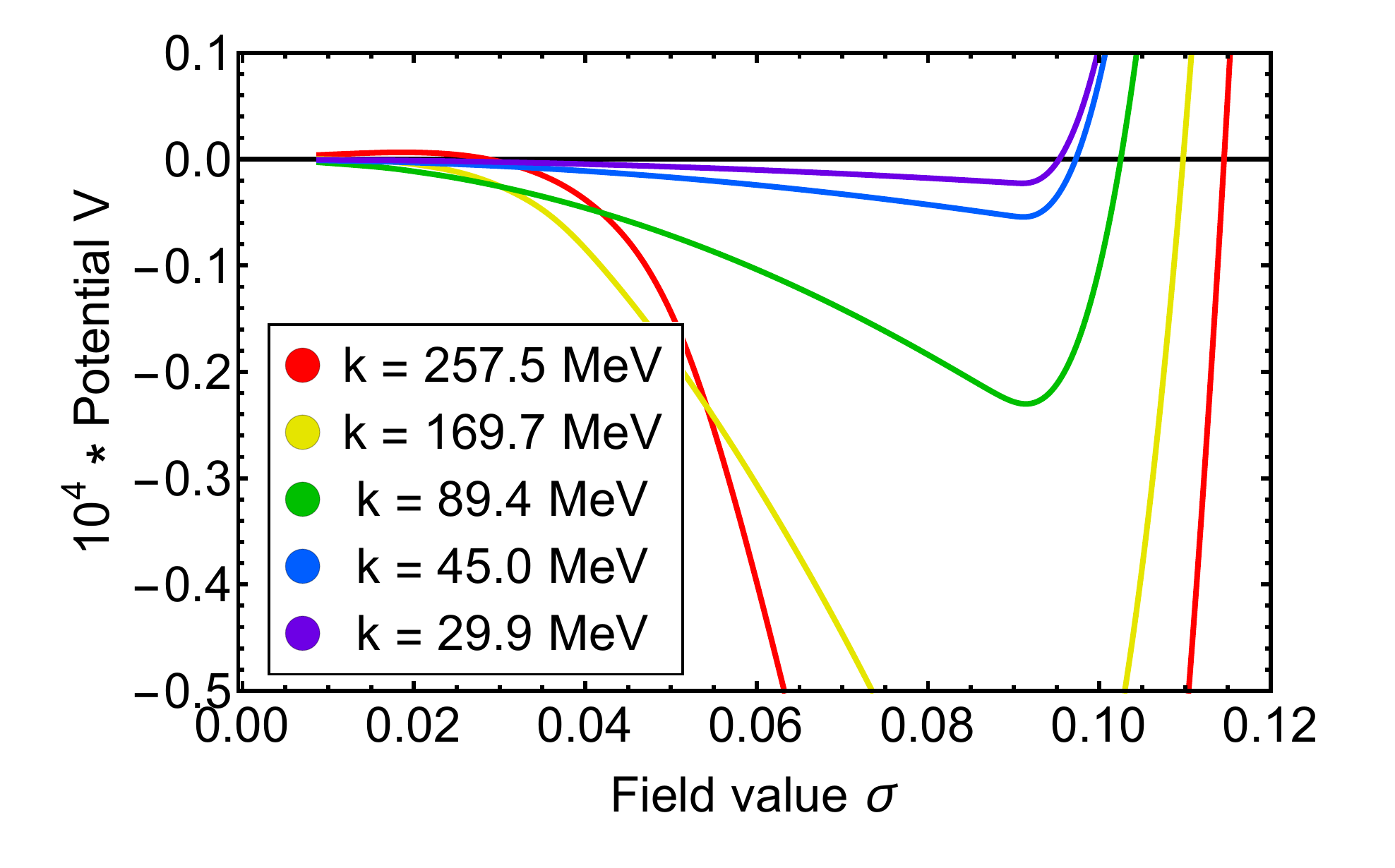}
		\subcaption{Potential in the broken phase.\hspace*{\fill} }
		\label{fig:si35}
	\end{minipage}%
	\hspace{1mm}
	\begin{minipage}[b]{.49\linewidth}
		\includegraphics[width=\linewidth]{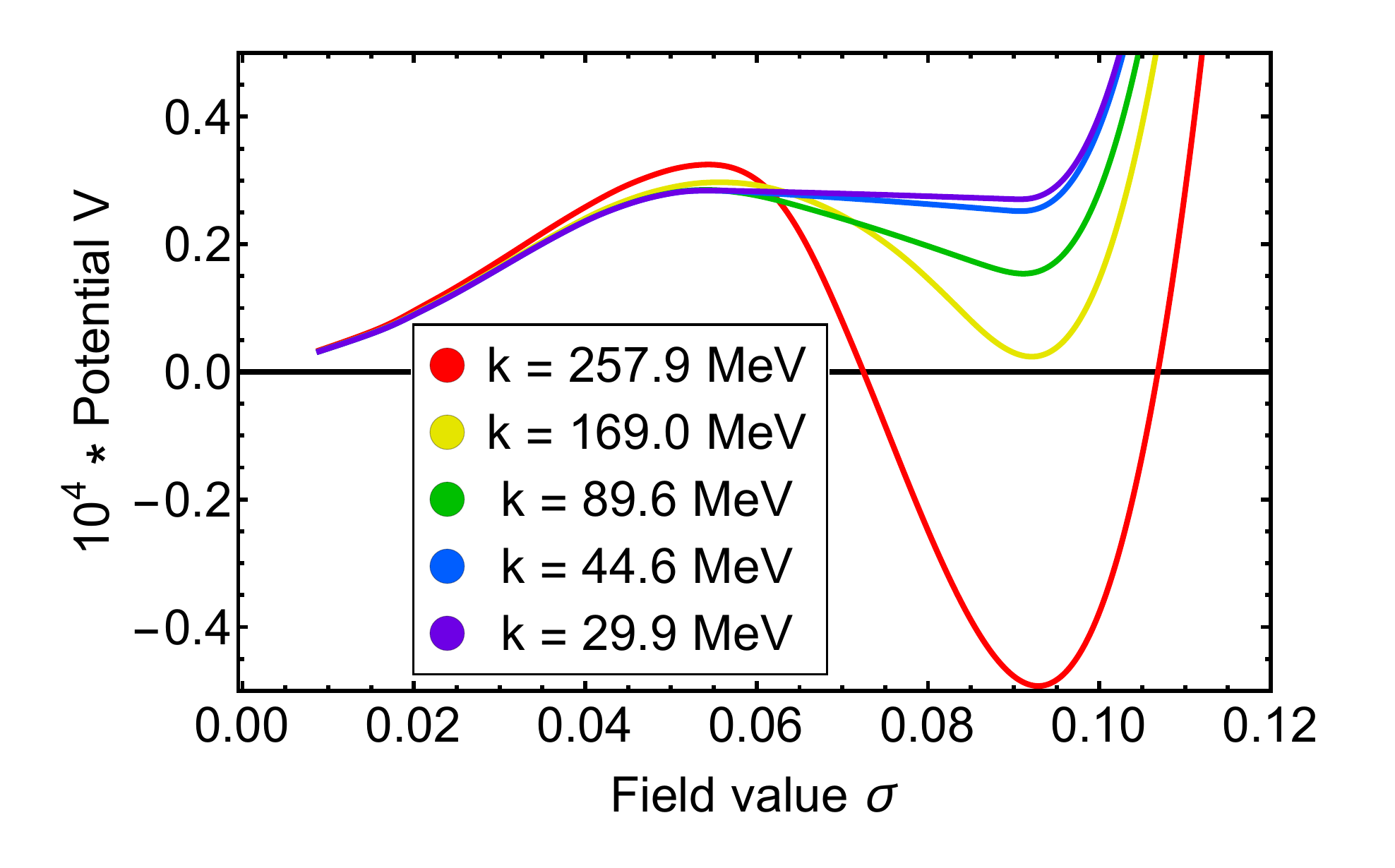}
		\subcaption{Potential in the symmetric phase.\hspace*{\fill} } \label{fig:si38}
	\end{minipage}	
	\caption{The effective potential $V_k(\rho)$ in the vicinity of the phase transition at $\mu = 0.30$ GeV (\Cref{fig:si35}) and $\mu = 0.32$ GeV (\Cref{fig:si38}) in the large $N_f$ limit with $N_\pi=3$.\hspace*{\fill}  }	
	\label{fig:shock1}
\end{figure*}
%

\section{Calculation of the Non-Conservative Numerical Flux}
\label{app:non_cons_flux}

The flow equation for the Yukawa coupling \labelcref{eq:hdrho} was reformulated in \Cref{app:flow} to suit the general form of the partial differential equations given in \labelcref{eq:pdeform} and contains a non-conservative flux term and a source term $s$, 
\begin{align*}
\label{eq:mq2cons2} 
	\partial_{t} w_{k} = A(u_k) \partial_{\rho} w_{k}+ s(u_{k} , w_{k} )
\, .
\end{align*}

The exact definition and derivation of the appearing terms is given in \Cref{app:thfkt}.

The non-conservative flux is computed using the integral derived by the jump condition in \Cref{app:numerics} in \labelcref{eq:noncons}. We chose a straight path across an interface, 
\begin{align*}
w(s) =&\, w^- + s (w^+ - w^-)  \,, \\[1ex]
u(s) =&\, u^- + s (u^+ - u^-)  \,.
\end{align*}
We note again that this is a path along the solutions $u$ and $w$ and not a path in the 'spatial' coordinate $\rho$. The straight path was chosen because it is often the simplest choice for the evaluation of the integral. In our case the expression simplifies so much that it can be evaluated analytically, due to the explicit form of the equations, where the non-conservative flux is given by
\begin{align}
A(u(s))= \partial_u g\left(u(s)\right)= \frac{1}{u^+ - u^-} \partial_s g\left(u(s)\right)\,.
\end{align}
This gives
\begin{align}\nonumber
\mathbf{D}(&\,u^+, u^-,w^+, w^-,\hat{\mathbf{n}}) \\[1ex]\nonumber
=&\frac{1}{2(u^+ - u^-)} \int_{0}^{1} \hat{\mathbf{n}} \frac{\partial g(u(s))}{\partial s} \frac{\partial w(s)}{\partial s} \ ds + C \mathbf{[[w]]} \\[1ex]
=&\frac{\hat{\mathbf{n}}}{2} \frac{g(u^+)-g(u^-)}{u^+ - u^-} \ (w^+ - w^-) + C \mathbf{[[w]]} \,,
\end{align}
where we used in the last equality that $\partial_s w$ is a constant expression. Instead, the constant $C$  is simply the absolute value of the jacobian matrix. 
\begin{equation}
C =  \int_0^1 |A(u(s)) \hat{\mathbf{n}} | \ ds \,.
\end{equation}
Often, it can be approximated as the maximal characteristic speed of the non conservative product.  Note that for constant $u$ across the interface $\frac{g(u^+)-g(u^-)}{u^+ - u^-} = A(u)$, such that we recover a conservative flux for constant $u$. There is a large set of paths across the interface that lead to the same value in the integral, due to the fact that $A$ can be written as a derivative of $u$. This hints at the possibility that there might be a conservative formulation for the system of equations. 

Since this formulation only allows flux boundary conditions for conservative fluxes a partial integration is performed, 
\begin{align}
A(u_k) \partial_\rho w_k = \partial_\rho (A(u_k) w_k) - w_k \partial_\rho A(u_k) \,.
\end{align}
We now have a conservative flux $A(u_k) w_k$ the proper in-/out-flow boundary conditions for $w_k$ and a very small non-conservative flux contribution $\mathbf{D'}$ accounting for jumps in $u_k$, 
\begin{align}
\mathbf{D'} = \mathbf{D} -[A(u_k^+) w_k^+ - A(u_k^-) w_k^-] \,.
\end{align}
This contribution is very small when $u_k$ is smooth and only contains small jumps across interfaces. It obviously vanishes at the outer boundary since there we have $u_k^+ =u_k^-$. Thanks to this formulation of the equation the maximal wave speed of the non conservative product is rather  small and can be neglected in practice. In the general case however the inclusion of this term is important especially if the  non conservative product are the only convective term in the equation, since it introduce the necessary numerical dissipation to make the numerical scheme stable.

\begin{figure*}[t]
	\centering
	\begin{minipage}[b]{0.49\linewidth}
		\includegraphics[width=\linewidth]{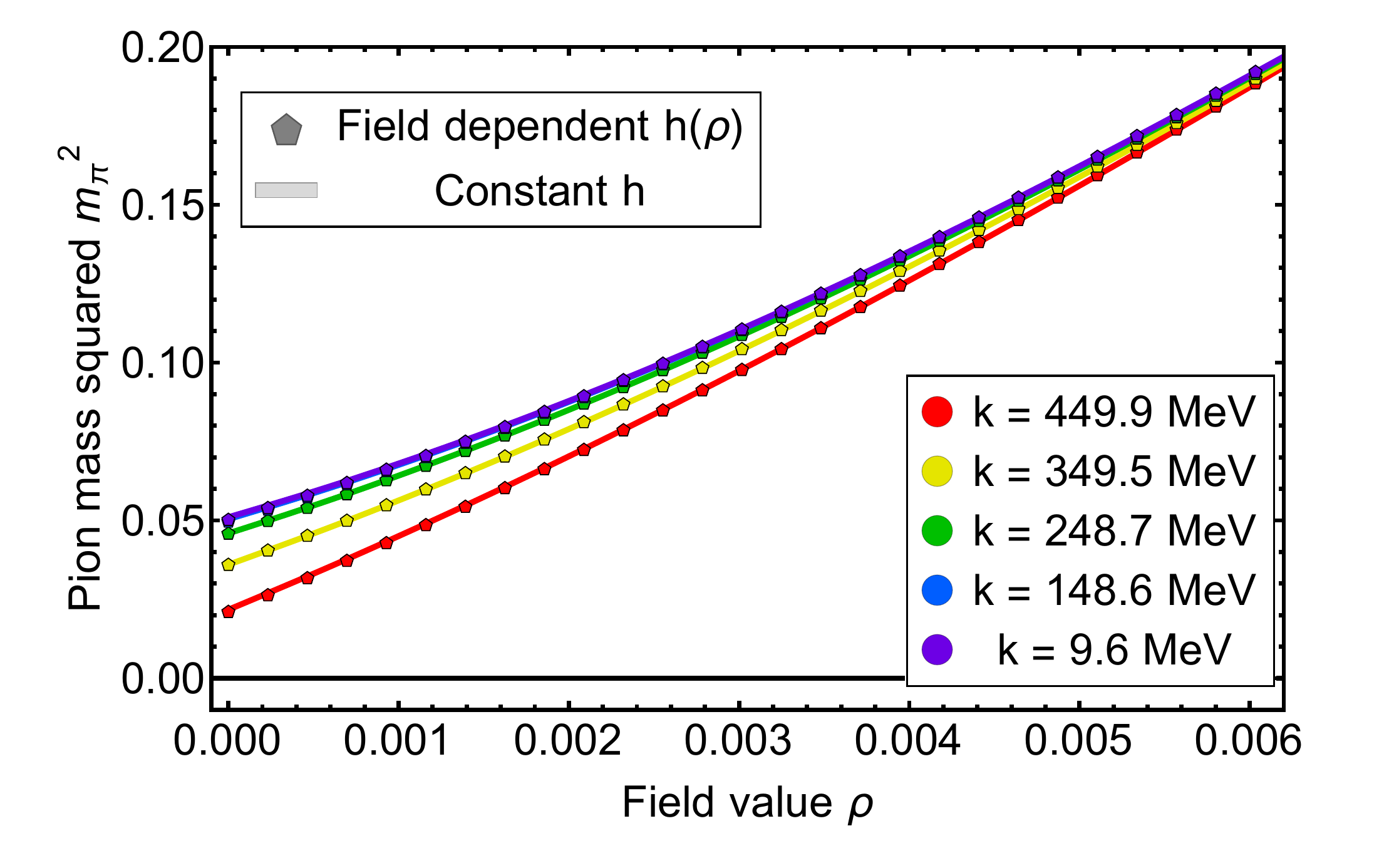}
		\subcaption{Pion mass at high temperatures ($T = 280$ MeV) and zero density.\hspace*{\fill} }
		\label{fig:ypt}
	\end{minipage}%
\hspace{1mm}
	\begin{minipage}[b]{0.49\linewidth}
		\includegraphics[width=\linewidth]{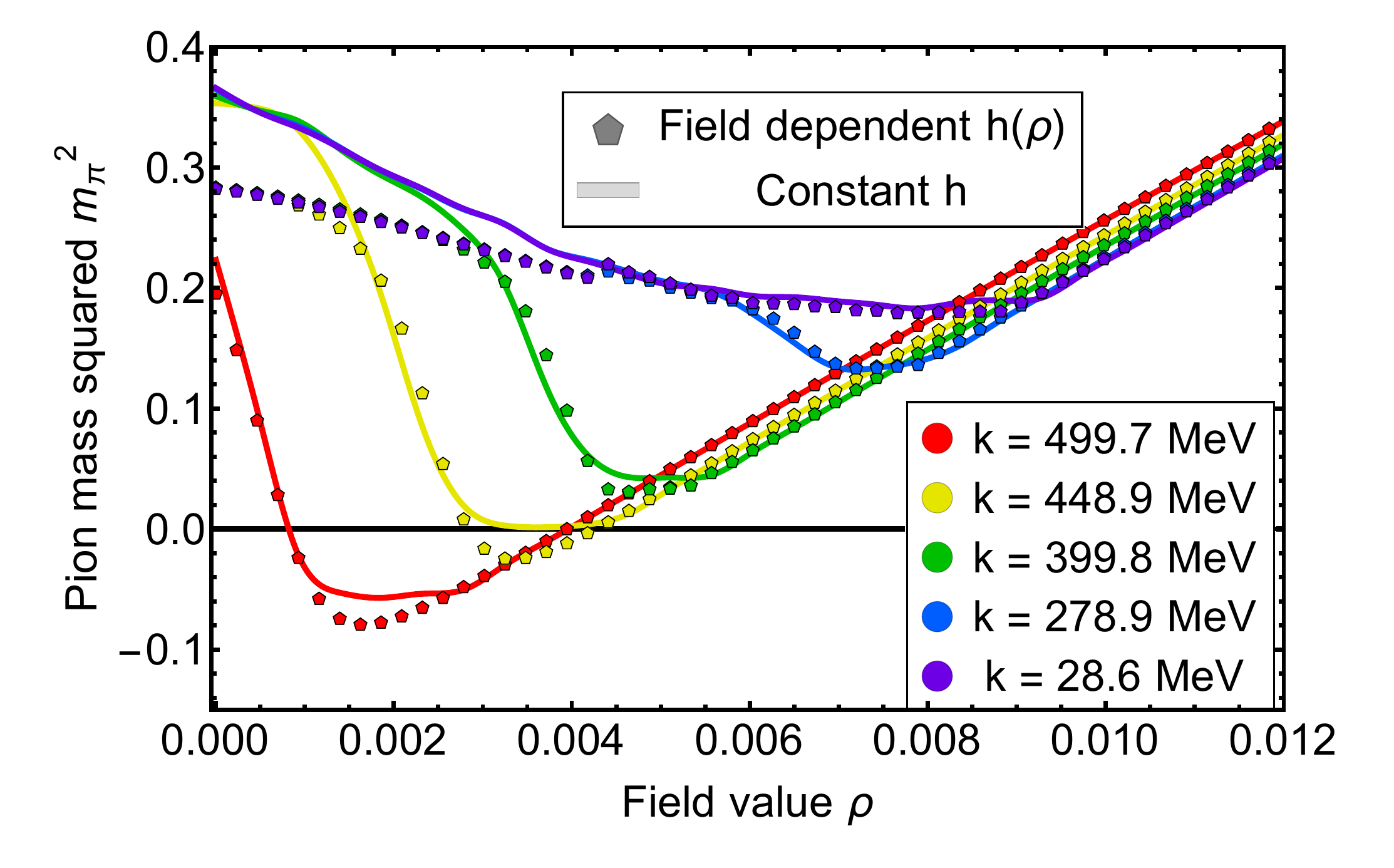}
		\subcaption{Pion mass at high densities ($\mu = 460$ MeV) and zero temperature.\hspace*{\fill} } \label{fig:ypc}
	\end{minipage}
	\begin{minipage}[b]{0.49\linewidth}
		\includegraphics[width=\linewidth]{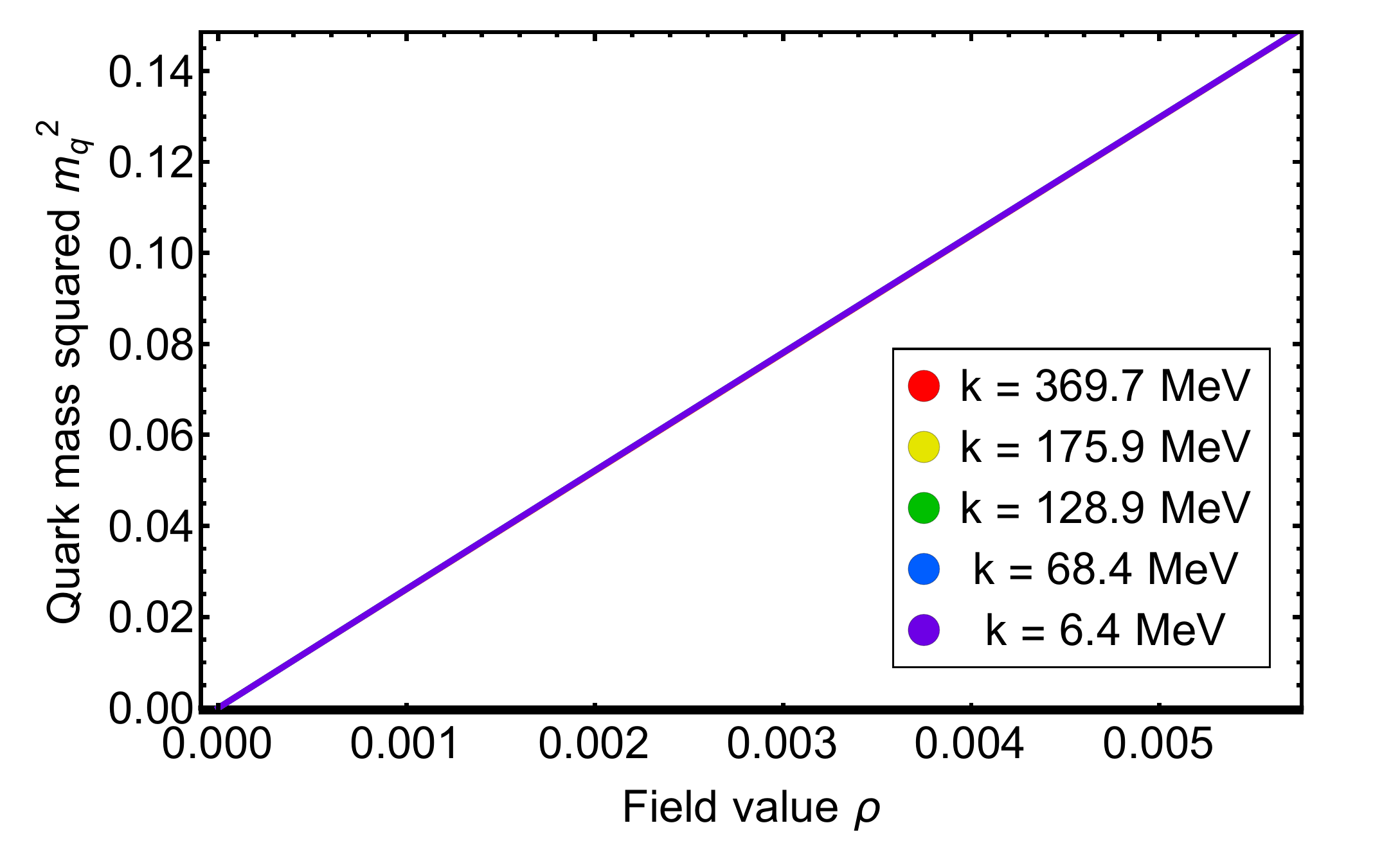}
		\subcaption{Quark mass at high temperatures ($T = 280$ MeV) and zero density.\hspace*{\fill} }
		\label{fig:yqt}
	\end{minipage}%
\hspace{1mm}
	\begin{minipage}[b]{0.49\linewidth}
		\includegraphics[width=\linewidth]{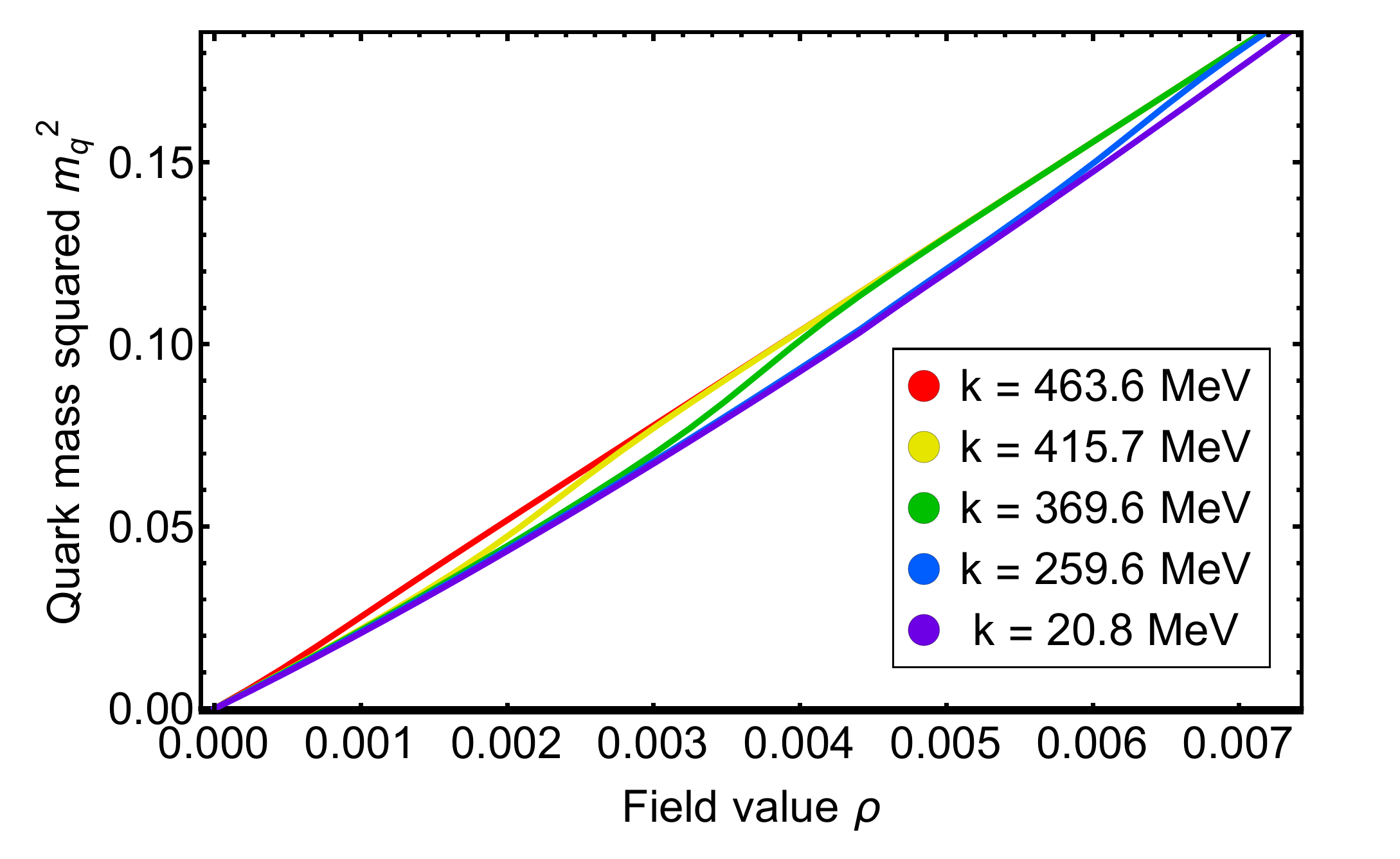}
		\subcaption{Quark mass at high densities ($\mu = 460$ MeV) and zero temperature.\hspace*{\fill} } \label{fig:yqc}
	\end{minipage}		
	\caption{RG-time evolution of the pion and quark masses in the broken symmetry phase. The computation was performed using $K=120$ cells and a local approximation order of $N_p =2$.\hspace*{\fill} }	
	\label{fig:yfinitetc}
\end{figure*}

\section{Shock Development}\label{app:shock}	
In this section we illustrate the dynamics of shock development in the large-N model.
In \Cref{sec:result} we distinguished two scenarios of shock development which are both illustrated in \Cref{fig:shock2}. \Cref{fig:s35} shows the pion mass/the potential derivative in the broken symmetry phase, where the shock eventually proceeds to move to unphysical values for $\rho$. This leads to the creation of a temporary maximum in the potential, depicted in \Cref{fig:si35}, which flattens out again for $k \to 0$. \Cref{fig:si35} shows the typical form of a potential in the broken symmetry phase with a degenerate global minimum at $k=0$. In \Cref{fig:s38} the shock freezes at some positive finite value. The temporary maximum still vanishes due to convexity restoration, however it remains at a positive value such that the potential in the symmetric phase has a unique global minimum at $\sigma_0 = 0$\,GeV.

\section{Field-dependent pion and quark masses}\label{app:resyuk}
Here we provide plots for the field dependent quark mass $m_q^2$ and pion mass $m_\pi^2$ for the case of high temperatures and high chemical potential in the symmetric phase. It can be seen from \Cref{fig:ypt} and \Cref{fig:ypc} that the RG-flow generates massive pions, similar to the computations with constant Yukawa coupling. The quark mass is expected to vanish in the symmetric phase as can be seen from \Cref{fig:yqc}.

\section{Regulators and threshold Functions} \label{app:thfkt}

In the present work we use the 3-dimensional flat or Litim regulator, see \cite{Litim:2002cf}. 
The flat bosonic regulator is,   
\begin{align}\label{eq:regphi} 
\hspace{-.2cm} R_\phi(p) =&\, p^2 \,r_\phi(x)\,,\quad r_\phi(x) = \,\left(\frac{1}{x}-1\right)\theta(1-x)\,,
\end{align}
with $x=p^2/k^2$, and the fermionic one reads 
\begin{align}
R_q=&\,  \pslash  \, r_q(x)\,, \quad \hspace{.2cm}  r_q(x) =  \left(\frac{1}{\sqrt{x}}-1\right)\theta(1-x)\,, 
\label{eq:regq}\end{align}
The threshold functions $l_1^{B/F,d}$ and $L^{(d)}_{n,m}$ in \labelcref{eq:Ah} and \labelcref{eq:Bh} are taken from \cite{Pawlowski:2014zaa} and shown for the sake of completeness for $Z_\phi= Z_q=1$, see \labelcref{eq:z}. 

The functions $l_0^{(B/F,d)}$ are deduced from bosonic/fermionic loops in $d$ dimensions, to wit, 
\begin{align}\label{eq:l0} \nonumber
l_0^{(B/F,d)}&(\hat{m}_{\phi,k}^2; T) \\[1ex] \nonumber
&= \frac{T}{2k}\sum_{n \in \mathbb{Z}} \int dx x^\frac{d-1}{2}\frac{\partial_t r_\phi(x) - r_\phi (x) }{\frac{\omega_n^2}{k^2} + x [1+r_\phi(x)]+m_{\phi,k}^2} \\[1ex]
&=\frac{2}{d-1} \frac{1}{\sqrt{1+\hat{m}_{\phi,k}^2}} \left[\frac{1}{2} + n_B(T,\hat{m}_{\phi,k}^2)\right] \,. 
\end{align}
In \labelcref{eq:l0}, $n_B$ is the Bose-Einstein distribution and $\hat{m}^2 = {m^2}/{k^2}$ the dimensionless masses. The threshold functions $l_n^{(B/F,d)}$ for diagrams with loops containing $n+1$ bosonic/fermionic propagator terms are obtained by taking a derivative with respect to $\hat{m}^2$, 
\begin{align}\label{eq:ln}
\partial_{\hat{m}^2} l_n^{(B/F,d)} (\hat{m}^2) = -(n + \delta_{n0}) l_{n+1}^{(B/F,d)} (\hat{m}^2)\,.
\end{align} 
Thus the function $l_1^{(B,4)}$ corresponds to the last term in Figure \ref{fig:h}, containing a bosonic regulator and two propagators. 

The function $L_{(1,1)}^{(4)}$ in \labelcref{eq:Bh} consists of the first two terms in Figure \ref{fig:h}. It can be obtained with a $\hat{m}^2_{q,k}$ derivative from $\mathcal{FB}_{(1,1)}$, that describes a loop with one fermionic and one bosonic propagator, 
\begin{align}\label{eq:FB}\nonumber
\mathcal{FB}_{(1,1)}&(\hat{m}^2_{q,k},\hat{m}^2_{\phi,k};T,\mu)\\[1ex]
=&\ \frac{T}{k} \mathrm{Re} \left[\sum_{n \in \mathbb{Z}}\frac{1}{\frac{\omega_n^2}{k^2} + x \left[1+r_\phi(x)\right]+\hat{m}_{\phi,k}^2}\right.\nonumber \\[1ex]\nonumber
&\left.\times \frac{1}{\frac{(\nu_n+i \mu)^2}{k^2} + x \left[1+r_q(x)\right]^2+m_{q,k}^2}\right]\\[1ex]\nonumber
=&\ \mathrm{Re}\left\{ \frac{1}{2\sqrt{1+\hat{m}^2_{\phi,k}}} \left[\frac{1}{2}+n_B(T,\hat{m}^2_{\phi,k})\right] \right.\\[1ex]\nonumber
&\times \left[\frac{1}{\hat{m}^2_{q,k}+1-\left(\frac{\mu}{k} - \frac{i \pi T}{k} - \sqrt{1+\hat{m}^2_{\phi,k}}\right)^2}\right. \\[1ex]\nonumber
&+\left.\frac{1}{\hat{m}^2_{q,k}+1-\left(\frac{\mu}{k} - \frac{i \pi T}{k} + \sqrt{1+\hat{m}^2_{\phi,k}}\right)^2}\right] \\[1ex]\nonumber
&-\frac{1}{2\sqrt{1+\hat{m}^2_{q,k}}}\left(n_F(T,-\mu,\hat{m}^2_{q,k})
\phantom{\frac{1}{\left(\sqrt{\frac{1}{1}}\right)} }
\right. \\[1ex]\nonumber
&-\left. \frac{1}{2}\frac{1}{\hat{m}^2_{\phi,k}+1-\left(\frac{\mu}{k} - \frac{i \pi T}{k} - \sqrt{1+\hat{m}^2_{q,k}}\right)^2}\right) \\[1ex]
&-\frac{1}{2\sqrt{1+\hat{m}^2_{q,k}}}\left(n_F(T,\mu,\hat{m}^2_{q,k}) 
\phantom{\frac{1}{\left(\sqrt{\frac{1}{1}}\right)} }
\right.\\[1ex]\nonumber
&-\left.\left.\frac{1}{2}\frac{1}{\hat{m}^2_{\phi,k}+1-\left(\frac{\mu}{k} - \frac{i \pi T}{k} + \sqrt{1+\hat{m}^2_{q,k}}\right)^2}\right)
\right\} \,,
\end{align} 
where
\begin{align}
n_F(T,\mu,\hat{m}^2_{q,k})= \frac{1}{\exp\left\{\frac{k}{T}\left( \sqrt{1+\hat{m}^2_{q,k}}+\frac{\mu}{k} \right)\right\}+1}\,.
\end{align} 
$L_{(1,1)}^{(4)}$ can now be generated by taking derivatives with respect to the fermionic mass $\hat{m}^2_{q,k}$ and the bosonic mass $\hat{m}^2_{\phi,k}$. These derivatives correspond to multiplying a fermionic/bosonic propagator to the loop in $ \mathcal{FB}_{(1,1)} $, 
\begin{align}\nonumber
\partial_{\hat{m}_{q,k}^2} \mathcal{FB}_{(m,n)}=&-m \mathcal{FB}_{(m+1,n)} \,, \\[1ex] \nonumber
\partial_{\hat{m}_{\phi,k}^2} \mathcal{FB}_{(m,n)}=&-n \mathcal{FB}_{(m,n+1)} \,.	
\end{align}
This leads us to 
\begin{align}\label{eq:L1}\nonumber
L_{(1,1)}^{(d)} =&\, \frac{T}{2k}\sum_{n \in \mathbb{Z}}\int dx x^{\frac{d-1}{2}} \left\{\frac{\partial_t r_\phi(x)}{\frac{\nu_n^2}{k^2} + x\left[1+r_q(x)\right]^2+m_{q,k}^2} \right.\\[1ex] \nonumber
& \hspace{1.4cm}\times
\left(\frac{1}{\frac{\omega_n^2}{k^2} + x\left[1+r_\phi(x)\right]+m_{\phi,k}^2}\right)^2  \\[2ex] \nonumber
&\hspace{1cm}+\frac{2 r_q (x)}{\frac{\omega_n^2}{k^2} + x\left[1+r_\phi(x)\right]+m_{\phi,k}^2} \\[3ex] \nonumber
&\hspace{1.4cm}\left.\times \left(\frac{1}{\frac{\nu_n^2}{k^2} + x\left[1+r_q(x)\right]^2+m_{q,k}^2}\right)^2\right\}\\[1ex]
=&\,
\frac{2}{d-1}\big( \mathcal{FB}_{(1,2)} +  \mathcal{FB}_{(2,1)}\big) \,.
\end{align} 
%


\endgroup

\bibliography{ref-lib} 

\end{document}